\documentclass[aps,pre,preprint,groupedaddress,showpacs]{revtex4}
\usepackage{amssymb}
\usepackage{amsfonts}
\usepackage{mathbbol}
\usepackage{stmaryrd}
\usepackage{graphicx}
\usepackage{url}
\usepackage{epstopdf}

\begin{document}
\urldef{\imageurl}\url{http://jfi.uchicago.edu/~tten/ChiralSedimentation/}

\title{Chiral sedimentation of extended objects in viscous media}

\author{Nathan W. Krapf}

\affiliation{University of Chicago}
\author{Thomas A.  Witten}
\affiliation{University of Chicago}
\author{Nathan C. Keim}
\affiliation{University of Chicago}

\date{\today}

\begin{abstract}
We study theoretically the chirality of a generic rigid object's sedimentation in a fluid under gravity in the low Reynolds number regime. We represent the object as a collection of small Stokes spheres or stokeslets, and the gravitational force as a constant point force applied at an arbitrary point of the object. For a generic configuration of stokeslets and forcing point, the motion takes a simple form in the nearly free draining limit where the stokeslet radius is arbitrarily small. In this case, the internal hydrodynamic interactions between stokeslets are weak, and the object follows a helical path while rotating at a constant angular velocity $\omega$ about a fixed axis. This $\omega$ is independent of initial orientation, and thus constitutes a chiral response for the object. Even though there can be no such chiral response in the absence of hydrodynamic interactions between the stokeslets, the angular velocity obtains a fixed, nonzero limit as the stokeslet radius approaches zero. We characterize empirically how $\omega$ depends on the placement of the stokeslets, concentrating on three-stokeslet objects with the external force applied far from the stokeslets. Objects with the largest $\omega$ are aligned along the forcing direction. In this case, the limiting $\omega$ varies as the inverse square of the minimum distance between stokeslets. We illustrate the prevalence of this robust chiral motion with experiments on small macroscopic objects of arbitrary shape.
\end{abstract}



\maketitle

\section{Introduction}
\label{sec:intro}

It is not unusual to see objects falling through water or air twisting as they sink. For example, a propeller-like maple seed will twirl as it falls from the tree. A consistent preference for twisting in a particular direction would constitute a chiral response of the object. Such a response must reflect some chirality in its shape, and the magnitude and nature of the twisting is evidently a consequence of well-known hydrodynamic laws. However, there is little fundamental understanding of what features of the shape control the magnitude of a chiral response.

In the past decade there has been a revival of interest in the tumbling motion exhibited by extended objects as they fall through air \cite{Belmonte:1998p12,Pesavento:2004p9}. These complex motions are of a different nature than what we study here. The objects under consideration have no intrinsic chirality, and interesting motions depend instead on significant Reynolds numbers, where the advection of momentum through the fluid is important. 

Aside from these, a few studies have examined the low Reynolds number sedimentation of different bodies. For a specific propeller-like design, Makino and Doi \cite{Makino:2003p4} showed that an ensemble of identical particles with different initial orientations will bunch together into a cylindrical shape oriented along the direction of the sedimenting force, whereas a similar group of achiral ellipsoids will drift apart. They have also made some headway in classifying the range of allowable motions for objects depending on whether or not they are skew or if there is an applied torque \cite{Doi:2005p8}. Gonzalez, Graf, and Maddocks \cite{Gonzalez:2004p10} have further explained some properties of the possible motions. We hope to improve on the parts of this understanding related to chiral objects.

Understanding the connection between shape and chiral motion would allow chiral sedimentation to be used as a characterization tool for objects of a supramolecular scale, such as colloidal particles and cells. Detecting the rotation of sedimenting bodies would give information not obtainable from other simple probes such as dynamic light scattering and intrinsic viscosity. These conventional measures sense only the hydrodynamic size of the objects, whereas rotation speed can sense the distinctive feature of a permanent chiral shape. Many biological structures have a strong chirality that is unrelated to propulsion. Examples include protein-DNA complexes \citep[Chapter 10]{Lodish:1995} and fibrils such as Actin \citep[Chapter 22]{Lodish:1995}, which are made of repeating subunits. Such objects must rotate as they sediment, and a knowledge of the connection between their shape and their rotation would be valuable.

We will show that chiral motions are natural to characterize when the hydrodynamic interactions between parts of the body are small. Thus, much of our study will be aimed at objects with this property, which we will term ``nearly free draining.'' Physical realizations of such objects can include thin, rod-like objects such as microtubules \citep[Chapter 23]{Lodish:1995}, bacterial flagella \citep[Chapter 23]{Lodish:1995}, or sickle cells \citep[Chapter 19]{Becker:2000}. As a concrete example, we can consider the propeller shape of Makino and Doi \cite{Makino:2003p4}, shown in Figure~\ref{propellerimage} below. For such an object of length about 10 microns in water, with a density of about 1 g/cm${}^3$, we predict rotational velocities on the order of 10 Hz. This should be noticeable, even when compared to the rotational diffusion coefficient, which for an object of this scale is only of order $10^{-4}$ Hz. Smaller objects on the scale of a micron or less will also have a noticeable effect if they sediment under a slightly larger force, as in a centrifuge.

In Section \ref{sec:propmat}, we discuss the equations of motion for our objects, and show how any inherent chirality must be encoded and expressed. 
In Section~\ref{sec:tumblezone} we introduce the ``tumble zone,'' a region in parameter space which determines whether or not a sedimenting object can exhibit ongoing tumbling behavior, and put a bound on its size. Following that, in Section \ref{sec:stokesletmodel} we review a stokeslet formalism for modeling rigid bodies, and show how to use this to calculate the internal hydrodynamic interactions needed in the equations of motion. In Section \ref{sec:nearlyfreedrain}, we find these interaction effects in the nearly free draining limit, where the interaction strength becomes small. In this limit, we find that the tumble zone becomes arbitrarily small, and that almost all objects will exhibit chiral sedimentation. Once this is established, in Section \ref{sec:chirality} we show how the chiral response behaves in certain limiting cases. Using a simple three-stokeslet body, we empirically examine how different aspects of shape affect our measurement of chirality. In Section \ref{sec:example} we show the results of numerical simulations. We check these numerical results against the analytic ones found in Section~\ref{sec:nearlyfreedrain}, and compare the typical motions of a random chiral body with both a more symmetric propeller shape and an achiral ellipsoid. Finally, we report the results of a simple experiment done on small macroscopic objects of arbitrary shape.

Throughout the next several sections, we refer to many different types of objects. To distinguish them, we use the following conventions: 3-vectors and unit 3-vectors will be denoted with arrows (e.g. $\vec v$) and hats ($\hat v$), respectively. The $3\times 3$ matrices that operate on them will use a blackboard bold font ($\mathbb M$). 6-vectors will be in italics with vector signs ($\vec{\mathcal V}$), and the $6\times 6$ matrices will be underlined ($\underline{\mathbb M}$). Large vectors composed of 3-vectors for each stokeslet will be bolded with vector signs ($\vec{\mathbf v}$), and the matrices that interact with them will be bolded with underlines ($\underline{\mathbf M}$).

\section{The propulsion matrix}
\label{sec:propmat}

In order to analyze the behavior of our sedimenting body, we take advantage of the fact that at low Reynolds numbers, the force and torque on a body are proportional to its velocity and angular velocity. Following Purcell \cite{Purcell:1977p2}, we collectively refer to these constants of proportionality as the propulsion matrix $\underline{\mathbb P}$. That is, we define extended force and velocity vectors $\vec{\mathcal F}\equiv(\vec F,\vec \tau)^T$ and $\vec{\mathcal V}\equiv(\vec V, \vec \omega)^T$, and write
\begin{equation}
\label{eqn:propmatdef}
\vec{\mathcal F}=\underline{\mathbb P} \vec{\mathcal V}
\end{equation}

As a consequence of the Onsager relation, and the requirement that the dissipated energy be positive, this propulsion matrix must be both symmetric and positive-definite \cite{Happel:1983}, so it can be written in block form as
\begin{equation}
\label{eqn:propmatsubmatrices}
\underline{\mathbb P}=\left(\begin{array}{cc}\mathbb K & \mathbb C^T\\ \mathbb C & \mathbb\Omega \end{array}\right)
\end{equation}
where $\mathbb K$ and $\mathbb \Omega$ are symmetric $3\times 3$ matrices which are also positive-definite. 

The propulsion matrix contains all of the information necessary to describe the dynamics of the object. Once it is known, an analysis of the motion can be carried out without reference to the specifics of an object's shape.

In order to specify a torque, $\underline{\mathbb P}$ must be computed about a specific point. Moving this point will change both $\mathbb C$ and $\mathbb \Omega$, though $\mathbb K$ will remain the same. Happel and Brenner \cite{Happel:1983} show how each of these individually transform under a change of coordinates. We arrive at equivalent results in a slightly different form. To begin, let $\Sigma$ and $\Sigma'$ represent two different inertial frames used to describe variables. In the following, primed variables will denote quantities viewed in the $\Sigma'$ basis, and unprimed ones will be those living in the $\Sigma$ basis. We then have propulsion equations for each of the frames: $\vec{\mathcal F}=\underline{\mathbb P}\vec{\mathcal V}$ and $\vec{\mathcal F}'=\underline{\mathbb P}'\vec{\mathcal V}'$.

It is easy to transform between coordinate systems that differ only by a rotation: if $\mathbb R$ is the rotation matrix that will take one set of axes to the other, then each subblock $\mathbb X$ of $\underline{\mathbb P}$ changes as $\mathbb X\rightarrow \mathbb R\mathbb X\mathbb R^{-1}$. Next we consider frames $\Sigma$ and $\Sigma'$ which differ only by location of the origin, and let $\vec R$ be the vector to $\Sigma'$'s origin. We now consider the effects of a force $\vec F$ and torque $\vec\tau$ applied at the origin of $\Sigma$. The body will feel the same net force and torque, and thus respond with the same motion, that it will if we pull at $\vec R$ with force $\vec F$ and supply a torque of $\vec \tau+(-\vec R)\times \vec F$. That is,
\begin{equation}
\vec{\mathcal F}'=\vec{\mathcal F}+\left(\begin{array}{c}0\\ (-\vec R)\times \vec F \end{array}\right)\equiv (\underline{\mathbb 1}+\underline{\mathbb B})\vec{\mathcal F},
\end{equation}
where the matrix $\underline{\mathbb B}$ is defined in block form by
\begin{equation}
\label{eqn:Bdef}
\underline{\mathbb B}=\left(\begin{array}{cc}0 & 0\\ -\llbracket\vec R\times\rrbracket & 0 \end{array}\right).
\end{equation}
Here we are using the notation that for any vector $\vec X$, $\llbracket \vec X\times\rrbracket$ is the antisymmetric $3\times 3$ matrix which satisfies $\llbracket \vec X\times\rrbracket \vec v=\vec X\times\vec v$ for all vectors $\vec v$. 

There is some extended velocity vector associated with this given force and torque, but it will be represented differently in $\Sigma$ and $\Sigma'$. The angular velocity must be the same in both systems, but a different linear velocity needs to be used. Using $\vec V'+\vec R\times\vec \omega=\vec V+0\times\vec \omega$, we can conclude
\begin{equation}
\label{eqn:extendedVtransf}
\vec{\mathcal V}'=\vec{\mathcal V}+\left(\begin{array}{c}-\vec R\times\vec \omega\\ 0 \end{array}\right)=(\underline{\mathbb 1}-\underline{\mathbb B}^T)\vec{\mathcal V}.
\end{equation}

We can now combine these two expressions to get a relationship between $\underline{\mathbb P}$ and $\underline{\mathbb P}'$:
\begin{eqnarray*}
\vec{\mathcal F}'&=&\underline{\mathbb P}'\vec{\mathcal V}'\\
(\underline{\mathbb 1}+\underline{\mathbb B})\vec{\mathcal F}&=&\underline{\mathbb P}'(\underline{\mathbb 1}-\underline{\mathbb B}^T)\vec{\mathcal V}\\
\underline{\mathbb P}\vec{\mathcal V}&=&(\underline{\mathbb 1}+\underline{\mathbb B})^{-1}\underline{\mathbb P}'(\underline{\mathbb 1}-\underline{\mathbb B}^T)\vec{\mathcal V}.
\end{eqnarray*}
Since $\vec{\mathcal V}$ is an arbitrary velocity, we can just write
\begin{equation}
\underline{\mathbb P}=(\underline{\mathbb 1}-\underline{\mathbb B})\underline{\mathbb P}'(\underline{\mathbb 1}-\underline{\mathbb B}^T)
\end{equation}
where we have used the fact that $(\underline{\mathbb 1}+\underline{\mathbb B})^{-1}=(\underline{\mathbb 1}-\underline{\mathbb B})$.

There is a unique point, termed the ``center of reaction,'' \cite{Happel:1983} about which the submatrix $\mathbb C$ is symmetric. For many objects with a high degree of symmetry, this often coincides with the centers of mass and buoyancy, but for a general case, these different points are not related.

For the sedimentation processes that concern us, it is often convenient to deal with the inverse of the propulsion matrix, known as the mobility matrix $\underline{\mathbb M}$. We will write it in block form as
\begin{equation}
\label{eqn:mobilmatsubmatrices}
\underline{\mathbb M}=\left(\begin{array}{cc} \mathbb A & \mathbb T^T\\ \mathbb T & \mathbb S \end{array}\right).
\end{equation}

The matrix $\mathbb A$, which we will call the alacrity matrix, gives the velocity response to an applied force. Our screw matrix $\mathbb S$ gives the angular velocity caused by supplying a torque, and the twist matrix $\mathbb T$ shows the coupling between angular velocity and force. 

Since $\underline{\mathbb P}$, $\mathbb K$, and $\mathbb\Omega$ are symmetric and positive definite, $\underline{\mathbb M}$, $\mathbb A$, and $\mathbb S$ must be as well. There is also a unique choice of origin for which $\mathbb T$ is symmetric, but it is in general different from the center of reaction. We will call this point the center of twisting. Furthermore, by inverting the transformation law for $\underline{\mathbb P}$, we can find how $\underline{\mathbb M}$ changes if the origin is moved:
\begin{equation}
\underline{\mathbb M}=(\underline{\mathbb 1}+\underline{\mathbb B}^T)\underline{\mathbb M}'(\underline{\mathbb 1}+\underline{\mathbb B}).
\end{equation}
From this, one can see that the screw matrix remains invariant ($\mathbb S=\mathbb S'$), and that the twist matrix changes simply as 
\begin{equation}
\label{eqn:twisttransform}
\mathbb T=\mathbb T'-\mathbb S\llbracket\vec R\times\rrbracket.
\end{equation}

Conveniently, for sedimentation processes the twist matrix alone captures both the chiral information and the dynamics of interest. Indeed, if we want our sedimenting object to show a preferred chirality, $\underline{\mathbb M}$ must not be invariant under inversions about the origin. $\mathbb A$ is necessarily invariant under this inversion, since both force and velocity transform as vectors. The same is true for the screw matrix, since both torque and angular velocity transform as pseudovectors. However, the twist matrix will reverse sign. Thus any chirality in the object must manifest itself through this twist matrix. As a simple example, we see that if the center of twisting is at the origin, then an object can only be chiral if the eigenvalues of $\mathbb T$ are not symmetric about 0.

The physical manifestation of chirality we are concerned with is the rotation of our object: at any time $t$, $\vec\omega(t)=\mathbb T(t)\vec F+\mathbb S(t)\vec\tau$. However, since sedimentation involves forces acting on the centers of both mass and buoyancy, with no supplied torques on either, it is possible to choose as our origin a point of zero torque on the object. In this case, we just have $\vec\omega(t)=\mathbb T(t)\vec F$, which allows us to restrict our attention to the twist matrix.

The twist matrix scales in a simple way with the overall size of the sedimenting object \cite{Happel:1983}. For a given object, the force $\vec F$ needed to produce a given $\vec\omega$ is proportional to the viscosity, $\eta$. Thus $\mathbb T$ is inversely proportional to $\eta$: $\mathbb T=\eta^{-1}\tilde{\mathbb T}$, with $\tilde{\mathbb T}$ independent of viscosity. Evidently $\tilde{\mathbb T}$ has dimensions of viscosity / (force $\times$ time), or (length)${}^{-2}$. With a fixed force, the rotation rate for an object enlarged by a factor $\alpha$ will thus be reduced by a factor of $\alpha^2$. Analogous reasoning shows that the velocity $\vec V$ is reduced by a factor $\alpha^3$. The object's translation for a given increment of rotation thus varies linearly with $\alpha$, and enlarging the object simply enlarges the path of its sedimenting motion by the same factor.

Given the twist matrix at time $t$, it can be found some small $\Delta t$ later by rotating $\mathbb T(t)$ by the angle $\omega(t)\Delta t$. Then
\[
\mathbb T(t+\Delta t)=\big(\mathbb 1-\Delta t\llbracket\vec\omega(t)\times\rrbracket\big)\mathbb T(t)\big(\mathbb 1+\Delta t\llbracket\vec\omega(t)\times\rrbracket\big).
\]
Eliminating terms of order $\Delta t^2$ gives
\[
\mathbb T(t+\Delta t)=\mathbb T(t)+\Delta t\big[\mathbb T(t),\llbracket\vec\omega(t)\times\rrbracket\big]
\]
Taking $\Delta t\rightarrow 0$ yields
\begin{equation}
\dot{\mathbb T}=\big[\mathbb T,\llbracket \mathbb T\vec F\times\rrbracket\big].
\end{equation}

The evolution of the other blocks of the mobility matrix evolve in a similar fashion: $\dot{\mathbb A}=[\mathbb A,\llbracket \mathbb T\vec F\times\rrbracket]$, and likewise for $\mathbb S$. 

This formalism, with fixed axes in the lab frame and a dynamical $\mathbb T$, is equivalent to the Euler equation formalism used by Gonzalez et. al. \cite{Gonzalez:2004p10}, which treats the body axes as fixed, and considers a dynamic force vector. We denote quantities in this body frame of reference using double-prime marks, $''$. At each instant the body frame rotates relative to the space frame with angular velocity $\vec\omega$, as noted above. Thus the space frame rotates with respect to the body frame at angular velocity $-\vec\omega$, and $\dot{\vec F''} = -\vec\omega\times\vec F''$. This $\vec\omega$, common to both frames, can be expressed equally in the body or space frame: $\vec\omega = \mathbb T\vec F = \mathbb T''\vec F''$.

Of particular interest are stationary states, in which the essential part of the motion is constant in time. In the body frame, a stationary state is one in which $\dot{\vec F''} = 0$. Since 
\begin{equation}
\label{eqn:fevolve}
\dot{\vec F''} = -\omega\times\vec F'' = -\mathbb T''\vec F''\times \vec F'',
\end{equation}
there is a stationary state if and only if $\vec F''$ is an eigenvector of $\mathbb T''$, with an eigenvalue that we denote as $\lambda$. Since $\vec F''$ is constant in time, $\vec \omega = \mathbb T''\vec F''$ must be as well. 

Because the twist matrix is $3\times 3$, it has either one or three real eigenvalues. In the case of a single real eigenvalue, the analysis above implies two fixed-point forces in opposite directions. The sign of the eigenvalue gives the chirality: a positive eigenvalue means that with the usual right-handed definition of angular velocity, the object twists as it descends in the direction of a right-handed screw. The chirality of the two fixed points is thus the same. However, the stability is not. The stability of the fixed point direction $\hat F_0''$ can be determined by considering the quantity $\hat F''\cdot \hat F_0''$. Its derivative $\dot{\hat F''}\cdot\hat F_0''$ determines whether $\hat F''$ moves toward or away from the fixed point with time. One may readily show \cite{Gonzalez:2004p10} that for a given $\hat F_0''$, the sign of this derivative is fixed for all $\hat F''\ne\pm\hat F_0''$. If this were not the case, then there would be some $\hat F''$ for which $\dot{\hat F''}\cdot\hat F_0=0$. To see that this is impossible, note that it either requires $\dot{\hat F''}\perp \hat F_0''$ or $\dot{\hat F''}=0$. Consider first the case where $\dot{\hat F''}\perp \hat F_0''$. Equation~\ref{eqn:fevolve} tells us that $\dot{\hat F''}\perp \hat F''$ and $\dot{\hat F''}\perp\mathbb T''\hat F''$. Since $\hat F_0''$ is the only eigenvector of $\mathbb T''$, this means that $\dot{\hat F''}$ has no component in any of the three independent directions $\hat F_0''$, $\hat F''$, and $\mathbb T''\hat F''$. This leaves us with the option that $\dot{\hat F''}=0$. However, this means that $\hat F''$ is a fixed point, which contradicts the assumption that $\hat F_0''$ is the only eigenvector of $\mathbb T''$. Thus $\dot{\hat F''}\cdot\hat F_0''\ne 0$ for all $\hat F''\ne\pm\hat F_0''$, meaning that $\dot{\hat F''}\cdot\hat F_0''$ has the same sign for all such $\hat F''$. If the sign is positive, then all $\hat F''$ move toward the $\hat F_0''$ axis and $\hat F_0''$ is then a globally stable fixed point. Evidently the opposite fixed point at $-\hat F_0''$ is globally unstable.

With three real eigenvectors, $\dot{\hat F''}\cdot \hat F_0''$ can vanish at points besides $\pm\hat F_0''$, so the global stability argument above is no longer valid. The simple chiral signature of the object is no longer present, and the motion becomes more complicated and depends on initial conditions \cite{Gonzalez:2004p10}. Happily, this case can be excluded for a large class of objects, as we show below.

\section{The tumble zone}
\label{sec:tumblezone}
Given a fixed shape for an object, we can choose the center of twisting as our origin. At this point the twist matrix is symmetric, meaning $\mathbb T$ must have three real eigenvalues. Next, keeping the object's shape fixed, we can explore the locus of points to which we can move the forcing point while still keeping all three eigenvalues real. Since, as shown above, global stability is not present at these forcing points, we call the region they form the ``tumble zone'' for that particular shape. We will show here that the volume of this tumble zone is always finite.

With any choice of origin, the screw matrix $\mathbb S$ is always symmetric with positive eigenvalues, as discussed in Section~\ref{sec:propmat}. We may then work in the basis where
\[
\mathbb S=\textrm{diag}(s_1,s_2,s_3).
\]
In this basis, we will move the forcing point to $\vec R_p$. From Equation~\ref{eqn:twisttransform}, this will give us a new twist matrix $\mathbb T = \mathbb T'-\mathbb S\llbracket\vec R_p\times\rrbracket$, where $\mathbb T'$ is the twist matrix computed about the center of twisting. We will show that if we choose $\vec R_p$ to be sufficiently large, then the new twist matrix about this origin must have only one real eigenvalue.

We can compute the discriminant $\Delta$ of the characteristic polynomial of our new $\mathbb T$. If the discriminant of a cubic equation is positive, then there is one real root and two complex conjugate ones. In this case, our twist matrix will have only one real eigenvalue. The discriminant is
\begin{eqnarray}
\Delta = 27\textrm{Det}^2(\mathbb T)-4\textrm{Det}(\mathbb T)\textrm{Tr}^3(\mathbb T)+9\textrm{Det}(\mathbb T)\textrm{Tr}(\mathbb T)\big(\textrm{Tr}^2(\mathbb T)-\textrm{Tr}(\mathbb T^2)\big)\nonumber\\
-\frac 14 \textrm{Tr}^2(\mathbb T)\big(\textrm{Tr}^2(\mathbb T)-\textrm{Tr}(\mathbb T^2)\big)^2+\frac 12\big(\textrm{Tr}^2(\mathbb T)-\textrm{Tr}(\mathbb T^2)\big)^3, \label{eqn:discriminant}
\end{eqnarray}
which is homogeneous of degree 6 in $\mathbb T$. 

The discriminant $\Delta$ is a sixth degree polynomial in $R_p$, so if the coefficient of the $R_p^6$ term is positive, we can be assured of getting $\Delta>0$ for any $R_p$ bigger than the largest root of this polynomial. Since $\Delta$ is homogeneous, there can be no powers of $\mathbb T'$ in the $R_p^6$ term. This leading term can thus be found from Equation~\ref{eqn:discriminant} by replacing $\mathbb T$ with $\mathbb S\llbracket\vec R_p\times\rrbracket$. Since $\mathbb S$ is symmetric and $\llbracket\vec R_p\times\rrbracket$ is antisymmetric with a zero eigenvalue, Det$(\mathbb S\llbracket\vec R_p\times\rrbracket)$ and Tr$(\mathbb S\llbracket\vec R_p\times\rrbracket)$ both vanish. Accordingly, the only term in Equation~\ref{eqn:discriminant} that can contribute in order $R_p^6$ is the last one:
\begin{equation}
\label{eqn:order6discriminant}
\Delta = -\frac 12 \Big[\textrm{Tr}[(\mathbb S\llbracket\vec R_p\times\rrbracket)^2]\Big]^3+\mathcal O(R_p^5)
\end{equation}
In terms of the eigenvalues $s_i$ and the coordinates $R_{p1}$, $R_{p2}$, and $R_{p3}$, this trace has the form
\[
\textrm{Tr}[(\mathbb S\llbracket\vec R_p\times\rrbracket)^2] = -2\big(R_{p1}^2s_2s_3+R_{p2}^2s_1s_3+R_{p3}^2s_1s_2\big).
\]
Since the $s_i$ are all positive, if we define $s_m = $ min$\{s_i\}$ then we can write
\begin{equation}
\label{eqn:order6discriminant2}
\Delta \ge 2s_m^6R_p^6+\mathcal O(R_p^5),
\end{equation}
whose leading term has a positive coefficient.

Thus outside a sphere of sufficient radius $R_p$ the discriminant is positive, there is a single real eigenvalue, and the motion converges to the globally stable motion discussed in Section~\ref{sec:propmat}.

\section{The stokeslet representation}
\label{sec:stokesletmodel}

The propulsion matrix for a body can sometimes be found analytically, and there are several known results for objects with various symmetries \cite{Happel:1983}. However, it can be more difficult to find when such symmetries are not present. We use the approach of Kirkwood and Riseman, as described by Meakin and Deutch\cite{Meakin:1987p915}, in which a sedimenting body is represented as a rigid collection of small beads known as stokeslets. Each stokeslet corresponds to a point source of drag, which exerts a force proportional to its velocity: $\vec F=-\gamma \vec v$, with drag coefficient $\gamma=6\pi\eta\rho$ proportional to the fluid viscosity $\eta$ and effective radius $\rho$ of the stokeslet. 

By arranging the stokeslets appropriately, the flow field from most objects can be recreated \cite{Carrasco:1999p19}. Thus, they form a simple way of modeling arbitrary bodies. This approach is used, for example, to model flagellar propulsion \cite{Higdon:1979p5}. Carrasco and de la Torre \cite{Carrasco:1999p19} investigate the effectiveness of different strategies for placing the stokeslets.

To create a propulsion matrix from a collection of stokeslets, one must take into account the change in fluid velocity past each stokeslet caused by the presence of the others. If one does not include these hydrodynamic interactions, then there can be no chiral effects in the sedimentation: the object will sink straight down, so all drag forces will be vertical in order to oppose it, and thus there can be no torque about the vertical axis. However, if we do include these interactions, the velocity at each stokeslet may be perturbed from the vertical, possibly causing a torque about that axis. This can make the object demonstrate chirality by spinning.

The tool we use is the Oseen equation, which gives the change in fluid velocity caused by one of these stokeslets. In the frame of a body with $n$ stokeslets, let $\vec{\mathbf v}$ be the $3n$ component vector containing the velocity of the fluid at the locations of each stokeslet, taking hydrodynamic interactions into account: $\vec{\mathbf v}=(\vec v^1,\vec v^2,\ldots,\vec v^n)^T$. We define $\vec{\mathbf F}$ to be the $3n$ component force vector acting on the stokeslets, and $\vec{\mathbf v}_e$ the external (undisturbed) velocity of the fluid at the location of each stokeslet. Then we can write the Oseen equation, 
\[
\vec{\mathbf v}=\vec{\mathbf v}_e+\underline{\mathbf L}\vec{\mathbf F},
\]
where $\underline{\mathbf L}$ is the Oseen tensor \cite{Witten:2004}. If we denote particle number by Greek letters, and cartesian coordinate by Roman letters, then for $\alpha\ne\beta$ we can write
\begin{equation}
\underline{\mathbf L}_{ij}^{\alpha\beta}=\frac{1}{8\pi\eta r^{\alpha\beta}}\Bigg(\delta_{ij}+\frac{(r_i^\alpha-r_i^\beta)(r_j^\alpha-r_j^\beta)}{(r^{\alpha\beta})^2}\Bigg)
\end{equation}
with $r_i^\alpha$ the $i$th coordinate of particle $\alpha$, and $r^{\alpha\beta}$ the distance between particles $\alpha$ and $\beta$. For $\alpha=\beta$ we should have 0, since an individual stokeslet cannot affect itself.

Let $\underline{\mathbf U}$ be the $3n\times 6$ matrix which relates the $3n$ dimensional $\vec{\mathbf v}_e$ and the extended velocity vector $\vec{\mathcal V}=(\vec V,\vec\omega)^T$: $\vec{\mathbf v}_e=\underline{\mathbf U}\vec{\mathcal V}$. Since the velocity of the fluid past each stokeslet is the opposite of the velocity at which the object is moving through the fluid, $\vec{\mathbf v}_e^\alpha=-\vec V-\vec\omega\times \vec r^\alpha$. Thus we can see that $\underline{\mathbf U}=(\underline{\mathbf U}^1,\ldots \underline{\mathbf U}^n)^T$, with
\[
\underline{\mathbf U}^\alpha = (-\mathbb 1,\llbracket\vec r^\alpha\times\rrbracket) = \left(\begin{array}{cccccc} -1 & 0 & 0 & 0 & -r_z^\alpha & r_y^\alpha\\ 0 & -1 & 0 & r_z^\alpha & 0 & -r_x^\alpha\\ 0 & 0 & -1 & -r_y^\alpha & r_x^\alpha & 0\end{array}\right).
\]
This $\underline{\mathbf U}$ matrix also has the property that ${\vec{\mathcal F}}=\underline{\mathbf U}^T{\vec{\mathbf F}}$.

If we define the $3n\times 3n$ matrix $\underline{\mathbf\Gamma}=$diag$(\gamma^1,\gamma^1,\gamma^1,\gamma^2,\gamma^2,\gamma^2,\ldots,\gamma^n,\gamma^n,\gamma^n)$, then $\vec{\mathbf F}=\underline{\mathbf\Gamma} \vec{\mathbf v}$, and we can rewrite the Oseen equation as
\begin{eqnarray*}
\vec{\mathbf v}&=&\vec{\mathbf v}_e+\underline{\mathbf L}\underline{\mathbf\Gamma} \vec{\mathbf v}\\
\vec{\mathbf v}&=&(\underline{\mathbf 1}-\underline{\mathbf L}\underline{\mathbf\Gamma})^{-1}\vec{\mathbf v}_e\\
\underline{\mathbf U}^T\underline{\mathbf\Gamma} \vec{\mathbf v}&=&\underline{\mathbf U}^T\underline{\mathbf\Gamma}(\underline{\mathbf 1}-\underline{\mathbf L}\underline{\mathbf\Gamma})^{-1}\vec{\mathbf v}_e\\
{\vec{\mathcal F}}&=&\underline{\mathbf U}^T\underline{\mathbf\Gamma}(\underline{\mathbf 1}-\underline{\mathbf L}\underline{\mathbf\Gamma})^{-1}\underline{\mathbf U}\vec{\mathcal V}
\end{eqnarray*}
But this is just our definition of the propulsion matrix:
\begin{equation}
\label{eqn:propmatfromstokeslets}
\underline{\mathbb P}=\underline{\mathbf U}^T\underline{\mathbf\Gamma}(\underline{\mathbf 1}-\underline{\mathbf L}\underline{\mathbf\Gamma})^{-1}\underline{\mathbf U}
\end{equation}

This result, which is a straightforward extension of the Kirkwood Riseman method explained in Reference \cite{Carrasco:1999p19}, shows that one can calculate $\underline{\mathbb P}$ from a matrix inversion.

\section{The nearly free-draining limit}
\label{sec:nearlyfreedrain}

\subsection{The propulsion matrix}
\label{sec:nearlyfreedrain:propmat}

When using the stokeslet model, we have an obvious mechanism by which we can model nearly free draining bodies: we simply take the stokeslet size (and thus the drag coefficient $\gamma$) close to zero. If we assume from now on that each stokeslet has the same effective radius, we can obtain perturbative expansions in this common $\gamma$, and write
\begin{eqnarray*}
\mathbb K&=&\mathbb K_0\gamma+\mathbb K_1\gamma^2+\ldots\\
\mathbb\Omega&=&\mathbb \Omega_0\gamma+\mathbb\Omega_1\gamma^2+\ldots\\
\mathbb C&=&\mathbb C_0\gamma+\mathbb C_1\gamma^2+\ldots
\end{eqnarray*}

To first order in $\gamma$, there are no hydrodynamic interactions, so we just have
\[
\vec{\mathbf v}=\vec{\mathbf v}_e
\]
when the body is not rotating. In this case, the total force on the object is just the sum of the individual forces acting on each stokeslet: $\vec F=\sum_{\alpha=1}^n(-\gamma \vec{\mathbf v}^\alpha)$, and $\vec{\mathbf v}^\alpha=-\vec V$ is the same for all $\alpha$. But $\vec F=(\mathbb K_0\gamma)\vec V$, so we get a $\mathbb K_0$ that is just the identity matrix times the number of stokeslets $n$:
\begin{equation}
\label{eqn:k0def}
(\mathbb K_0)_{ij}=n\delta_{ij}
\end{equation} 

For the coupling tensor $\mathbb C$, we have
\begin{eqnarray}
(\mathbb C_0 \gamma)\vec V&=&\vec\tau=\sum_{\alpha=1}^n\vec r^\alpha\times(\gamma\vec V)\nonumber\\
\mathbb C_0&=&\sum_{\alpha=1}^n\llbracket\vec r^\alpha\times\rrbracket,
\end{eqnarray}
which is completely antisymmetric. If the origin is at the mean stokeslet position $\vec r^c = \frac 1n\sum_\alpha \vec r^\alpha$, then $\mathbb C_0=0$.

Finally, when the body is rotating without translating,
\[
(\mathbb\Omega_0\gamma)\vec\omega=\vec\tau=\sum_{\alpha=1}^n\vec r^\alpha\times(\gamma \vec{\mathbf v}^\alpha)=\gamma\sum_{\alpha=1}^n\vec r^\alpha\times(\vec\omega\times \vec r^\alpha)=\gamma\sum_{\alpha=1}^n\big((r^\alpha)^2-\vec r^\alpha \vec r^\alpha\cdot\big)\vec\omega,
\]
so $\mathbb\Omega_0$ is an inertia tensor:
\begin{equation}
\label{eqn:omega0def}
(\mathbb\Omega_0)_{ij}=\sum_{\alpha=1}^n\big((r^\alpha)^2\delta_{ij}-r^\alpha_ir^\alpha_j\big).
\end{equation}

To second order, hydrodynamic interactions become important:
\[
\vec{\mathbf v}=\vec{\mathbf v}_e+\underline{\mathbf L}(\gamma \vec{\mathbf v}_e).
\]

Using our expression for the Oseen tensor gives
\begin{eqnarray*}
(\mathbb K_1)_{ij}&=&\frac{1}{8\pi\eta}\sum_{\alpha,\beta\ne \alpha}\Big[\frac{\delta_{ij}}{r^{\alpha\beta}}+\frac{r^\alpha_i-r_i^\beta}{(r^{\alpha\beta})^3}(r^\alpha_j-r^\beta_j)\Big]\\
(\mathbb C_1)_{ij}&=&\frac{1}{8\pi\eta}\sum_{\alpha,\beta\ne\alpha}\Big[\frac{\llbracket\vec r^\alpha\times\rrbracket_{ij}}{r^{\alpha\beta}}-\frac{(\vec r^\alpha\times \vec r^\beta)_i}{(r^{\alpha\beta})^3}(r^\alpha_j-r^\beta_j)\Big]\\
(\mathbb \Omega_1)_{ij}&=&\frac{1}{8\pi\eta}\sum_{\alpha,\beta\ne\alpha}\Big[\frac{(\vec r^\alpha\cdot\vec r^\beta)\delta_{ij}-r^\alpha_ir^\alpha_j}{r^{\alpha\beta}}+\frac{(\vec r^\alpha\times \vec r^\beta)_i(\vec r_\alpha\times \vec r^\beta)_j}{(r^{\alpha\beta})^3}\Big]
\end{eqnarray*}

Since there is no term in the propulsion matrix which is zeroth order in $\gamma$, the expansion for $\mathbb T$ has the form
\[
\mathbb T=\mathbb T_0\gamma^{-1}+\mathbb T_1+\mathbb T_2\gamma+\ldots
\]
and will diverge as the effective stokeslet size approaches zero. Fortunately, the eigenvalues of $\mathbb T$ will not end up diverging as well.

To see this, we will compute $\underline{\mathbb M}$ by inverting $\underline{\mathbb P}$ in block form. We can identify the screw matrix $\mathbb S$ as the inverse of the Schur's complement of $\mathbb K$, giving $\mathbb S=(\mathbb \Omega-\mathbb C\mathbb K^{-1}\mathbb C^T)^{-1}$, and then $\mathbb T=-\mathbb S\mathbb C\mathbb K^{-1}$. Likewise, the alacrity matrix $\mathbb A$ is the inverse of the Schur's complement of $\mathbb\Omega$.

When we neglect internal hydrodynamic interactions, the twist matrix then becomes
\begin{equation}
\mathbb T_0 = -(\mathbb\Omega_0 - \mathbb C_0\mathbb K_0^{-1}\mathbb C_0^T)^{-1}\mathbb C_0\mathbb K_0^{-1}.
\end{equation}
As discussed above, there can be no twisting due to an applied force unless there are hydrodynamic interactions. This means that the centers of twisting and reaction are the same here, and that at this point, $\mathbb C_0 = \mathbb T_0 = 0$. As noted above, this point is also the mean stokeslet position $\vec r^c$. Since the twist matrix vanishes here, we can see that $\mathbb T_0$ will always have a null vector, regardless of where the origin is: to move the origin from the center of reaction to a position $\vec R$ corresponds to changing the twist matrix to $\mathbb T_0' = 0 -\mathbb S_0\llbracket \vec R\times\rrbracket$. But then $\mathbb T_0' \vec R = -\mathbb S_0\llbracket \vec R\times\rrbracket\vec R = 0$, so $\mathbb T_0'$ still has at least one eigenvector, $\vec R$, with a corresponding eigenvalue of zero. Thus as $\gamma$ decreases, one eigenvalue of $\mathbb T_0\gamma^{-1}$ remains zero, though the other two may become large and complex.

When small hydrodynamic effects are added, the twist matrix expands to first order as 
\[
\mathbb T = \mathbb T_0\gamma^{-1}+\mathbb T_1.
\]
Because $\gamma$ is small, $\mathbb T_1$ makes a negligible correction to the $\mathbb T_0\gamma^{-1}$ term, except in the null space of $\mathbb T_0$. Here, $\mathbb T\vec R = \mathbb T_1\vec R$, which is independent of $\gamma$. To this order, the axis of spin is then $\vec R$, the vector from the average stokeslet position $\vec r^c$ to the forcing point.

Since some eigenvalues of $\mathbb T_0$ can be complex, we cannot diagonalize it using real eigenvectors. However, we can put $\mathbb T_0$ into Jordan canonical form using a basis of the form $\{\vec v_1$, $\vec v_2$, $\vec R\}$. If we let $\vec R_d^T$ denote the dual of $\vec R$, satisfying $\vec R_d^T\vec R=1$ and $\vec R_d^Tv_i=0$, then a real eigenvalue of $\mathbb T$ to this order will be 
\[
\lambda = \vec R_d^T(\mathbb T_0/\gamma+\mathbb T_1)\vec R=\vec R_d^T\mathbb T_1\vec R
\]
which is independent of $\gamma$.

As noted above, the chiral response depends on hydrodynamic interactions between parts of the object. These interactions go to zero with the drag coefficient $\gamma$. Thus it is natural to anticipate that the angular velocity of the object should vanish with $\gamma$. Remarkably, this is not the case: we have just seen that a real eigenvalue of the twist matrix, and thus the angular velocity, reaches a non-zero limit as $\gamma\rightarrow 0$. In this sense, there is a qualitative difference between the nearly free draining state and the perfectly free draining state. The difference may be understood through the propulsion matrix, which gives the force and torque in terms of the velocity $\vec V$ and angular velocity $\vec\omega$, and is regular as $\gamma\rightarrow 0$. Both the amount of torque for a given $\vec V$ and no $\vec\omega$ and the amount of torque for a given $\vec\omega$ with no $\vec V$ are proportional to $\gamma$. With sedimentation, there is no net torque on the object, so we can find our $\vec\omega$ for a given $\vec V$ by the requirement that the torque vanishes. If $\gamma$ is then reduced, both sources of torque are reduced in proportion, and the total torque remains zero with no change in $\vec\omega$. Thus $\vec\omega$ has no tendency to vanish with $\gamma$. 

\subsection{The tumble zone}
\label{sec:nearlyfreedrain:tumblezone}

In Section~\ref{sec:tumblezone}, we showed that the tumble zone had finite volume. The size and shape of this volume depend on the drag coefficient $\gamma$. We will now show that the volume of the tumble zone goes to zero at least as fast as $\gamma^3$. Thus for sufficiently small $\gamma$, any collection of stokeslets taken about any origin with no special symmetries will fall outside of the tumble zone, and must thus have simple fixed point chiral sedimentation.

We use an argument similar to that in Section~\ref{sec:tumblezone}, but choose the forcing point to be of the form $\vec R_p=\gamma Q\hat R_p$, where $Q$ is independent of $\gamma$. About the center of twisting, we can write
\begin{eqnarray*}
\mathbb T&=&\mathbb T_1\\
\mathbb S&=&\mathbb S_0\gamma^{-1}+\mathbb S_1
\end{eqnarray*}
since in the low $\gamma$ limit, $\mathbb T_0=0$. Then
\begin{equation}
\label{eqn:lowgammatwisttransform}
\mathbb T = \mathbb T_1 - (\mathbb S_0/\gamma+\mathbb S_1)\llbracket \vec R_p\times\rrbracket = \mathbb T_1-Q(\mathbb S_0+\mathbb S_1\gamma)\llbracket\hat R_p\times\rrbracket
\end{equation}
which has a part of order $\gamma^0$ and a correction of order $\gamma^1$. The resulting discriminant $\Delta$ for the characteristic polynomial of $\mathbb T$ can be computed from Equation~\ref{eqn:discriminant} as in Section~\ref{sec:tumblezone}, but replacing $\mathbb S\llbracket\vec R_p\times\rrbracket$ with $Q\mathbb S_0\llbracket\hat R_p\times\rrbracket+\mathcal O(\gamma Q)$. Using this substitution, we obtain a discriminant similar to Equation~\ref{eqn:order6discriminant}:
\begin{equation}
\Delta = -\frac 12 \Big[\textrm{Tr}[(Q\mathbb S_0\llbracket\hat R_p\times\rrbracket)^2]\Big]^3+\mathcal O(Q^5)+\mathcal O(Q^6\gamma)
\end{equation}
Letting $s_{0m}$ be the smallest of the eigenvalues of $\mathbb S_0$ gives the bound
\begin{equation}
\label{eqn:lowgammadeltabound}
\Delta\ge 2s_{0m}^6Q^6\gamma^0+\mathcal O(Q^6\gamma)+\mathcal O(Q^5),
\end{equation}
except in the unphysical case that the stokeslets are perfectly collinear. In this case, one of the eigenvalues of $\mathbb S_0$ is zero, and taking $\vec R_p$ perpendicular to this direction will make the $Q^6\gamma^0$ term vanish.

Since $\gamma$ is small, the main contribution to the coefficient of the $Q^6$ term is from the $\gamma^0$ part, which from Equation~\ref{eqn:lowgammadeltabound} is positive. For sufficiently large $Q$, we can then be assured that $\Delta>0$, giving one real eigenvalue for $\mathbb T$.

Thus we see that in the nearly free draining limit, the tumble zone can be fit inside of a sphere whose radius is proportional to the drag coefficient $\gamma$. As $\gamma\rightarrow 0$, the tumble zone then must become vanishingly small. Unless the sedimenting object has the special property that its forcing point is exactly at the center of twist, we will thus get only one real eigenvalue for the twist matrix. We then expect globally stable chiral motion as it sediments.

\section{Chirality}
\label{sec:chirality}

The globally stable motion expected for nearly free draining sedimenting objects lends itself naturally to defining a chirality. If we denote the real eigenvalue of $\mathbb T$ by $\lambda$, then $\lambda=\omega/F$ for $F$ the magnitude of the applied force and $\omega$ the constant angular velocity. We can try to use this $\lambda$ as a measure of the chirality. Conveniently, $\lambda$ is independent of $\gamma$ for nearly free draining objects, so we only need to know the shape of the object and the forcing point, and are not obliged to worry about the precise stokeslet strength.

Unfortunately, if we try to use this measure to look for a ``most chiral'' object, we will be sorely disappointed: for a fixed $\gamma$, $\lambda$ diverges as the stokeslets become collinear. In this rather unphysical case, the eigenvalue of $\mathbb\Omega$ corresponding to rotations about the line of stokeslets will become zero, making $\underline{\mathbb P}$ non-invertible, and our expression for $\mathbb T$, which depends on $\mathbb\Omega^{-1}$, diverge. 

\subsection{The distant forcing point limit}
\label{sec:chirality:distant}

In order to characterize the divergence of $\lambda$ we may simplify the analysis by considering the limit where the forcing point is far away from the stokeslets. This is a convenient choice because as long as the distance $R_p$ from the center of reaction to the origin is large, $\lambda$ is actually independent of the precise value of $R_p$. This is true for any object, and does not depend on the approximation of small $\gamma$ used in Section~\ref{sec:nearlyfreedrain}.

To prove this assertion, we will first assume that we know the twist matrix around the center of twist. This choice of origin is somewhat arbitrary - any point close to the stokeslets will do. Once we have this $\mathbb T$, we will move the origin to the point $\vec R_p$, where according to Equation~\ref{eqn:twisttransform} the new twist matrix is given by $\mathbb T'=\mathbb T-\mathbb S\llbracket\vec R_p\times\rrbracket$. 

One of the eigenvalues of $S\llbracket\vec R_p\times\rrbracket$ is zero. Since Tr$(S\llbracket\vec R_p\times\rrbracket)=0$ and Tr$[(S\llbracket\vec R_p\times\rrbracket)^2]<0$, the two nonzero eigenvalues must be imaginary.

Next we will choose the basis, not necessarily orthogonal, which puts $\mathbb S\llbracket \vec R_p\times\rrbracket$ into Jordan canonical form. Here,
\[
\mathbb S\llbracket \vec R_p\times\rrbracket = \left(\begin{array}{ccc} r & 1 & 0\\ 0 & r & 0\\ 0 & 0 & 0 \end{array}\right)
\]
where $r$ is a generalized eigenvalue proportional to the pulling distance $R_p$. We will define the basis $\{|0\rangle$, $|1\rangle$, $|2\rangle\}$ by $\mathbb S\llbracket \vec R_p\times\rrbracket|0\rangle=0$, $\mathbb S\llbracket \vec R_p\times\rrbracket|1\rangle = r|1\rangle$, and $\mathbb S\llbracket \vec R_p\times\rrbracket|2\rangle=r|2\rangle+|1\rangle$. We will also form the dual basis $\{\langle 0|$, $\langle 1|$, $\langle 2|\}$, which satisfies $\langle i|j\rangle=\delta_{ij}$. 

Our goal is to find the real eigenvalue of $\mathbb T'=\mathbb T-\mathbb S\llbracket \vec R_p\times\rrbracket$. Since $R_p$ is large, $\mathbb T$ serves as a small perturbation of the $\mathbb S\llbracket \vec R_p\times\rrbracket$ matrix. The real eigenvalue $\lambda$ must then be a perturbation of the single real eigenvalue of $\mathbb S\llbracket \vec R_p\times\rrbracket$, namely zero. We will express its corresponding eigenvector as $|v\rangle = |0\rangle+\epsilon_1|1\rangle+\epsilon_2|2\rangle$, choosing to scale it so that the coefficient of $|0\rangle$ is 1, and $\epsilon_i\ll 1$. With this expansion,
\begin{eqnarray*}
\mathbb T'|v\rangle & = & \lambda |v\rangle\\
\mathbb T(|0\rangle+\epsilon_1|1\rangle+\epsilon_2|2\rangle)-\mathbb S\llbracket \vec R_p\times\rrbracket(|0\rangle+\epsilon_1|1\rangle+\epsilon_2|2\rangle)&=&\lambda(|0\rangle+\epsilon_1|1\rangle+\epsilon_2|2\rangle)\\
\mathbb T|0\rangle+\epsilon_1\mathbb T|1\rangle+\epsilon_2\mathbb T|2\rangle-\epsilon_1r|1\rangle-\epsilon_2r|2\rangle-\epsilon_2|1\rangle&=&\lambda(|0\rangle+\epsilon_1|1\rangle+\epsilon_2|2\rangle)
\end{eqnarray*}
The $\epsilon_i \mathbb T$ terms must be small by comparison with the $\epsilon_i r$ terms, so we can drop them. Now applying $\langle 0|$ to both sides gives
\begin{equation}
\lambda = \langle 0|\mathbb T|0\rangle,
\end{equation}
which is independent of the distance $R_p$. 

\subsection{Shape dependence of the chiral response}
\label{sec:chirality:shape}

Here we determine how the chiral sedimentation coefficient $\lambda$ depends on the locations of the stokeslets in the nearly free draining limit, in the case of distant forcing point. Even though $\lambda$ is independent of the distance to the forcing point in this limit, it can still depend on the orientation of the object relative to the pulling direction. We thus distinguish the coordinates of the stokeslets parallel and perpendicular to this forcing direction, denoted as $\hat z$. We first note that our system has no distinguished origin, so $\lambda$ can depend only on the distances between the stokeslets. Accordingly, we measure stokeslet positions relative to their center,
\[
\vec r^{c}=\frac 1n \sum_{\alpha=1}^n \vec r^\alpha.
\]
In terms of this, we define parallel and transverse radii of gyration, given by
\[
R_{\parallel}^2 = \frac 1n\sum_{\alpha=1}^n(r_z^\alpha-r_z^{c})^2
\]
and
\[
R_{\perp}^2 = \frac 1n\sum_{\alpha=1}^n|\vec r_\perp^\alpha - \vec r^{c}_\perp|^2.
\]
The total radius of gyration is then $R_g^2=R_{\parallel}^2+R_\perp^2$.

We use four parameters to characterize the distribution of stokeslets. The overall size can be expressed in terms of the radii of gyration given above. In addition, we use a length $Z$ defined below to characterize inhomogeneity in longitudinal position, and a dimensionless quantity $\Delta$ to characterize anisotropy in the transverse plane.

To simplify matters, we will focus on configurations with the fewest number of stokeslets required to make a chiral response possible. Since the object as a whole also includes a forcing point, we only need three stokeslets to guarantee a non-planar configuration. In such cases, with a distant forcing point, there are nine coordinates which can specify shape. However, $\lambda$ is independent of translation and of rotation around the pulling axis, so only five coordinates are potentially significant. We next show that the four parameters named above appear to suffice.

\begin{figure*}
\begin{tabular}{ccc}
\resizebox{50mm}{!}{\includegraphics{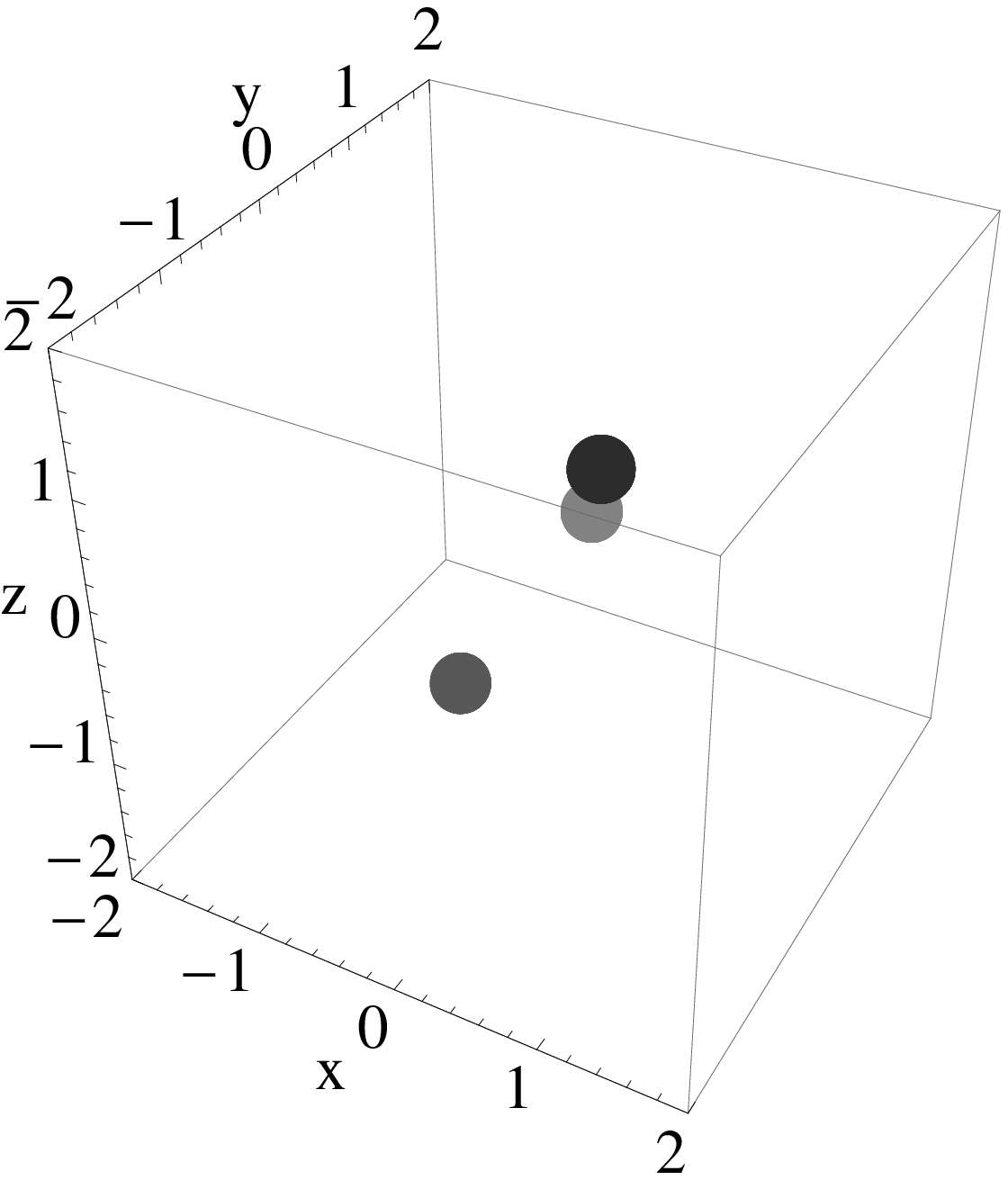}} &
\resizebox{50mm}{!}{\includegraphics{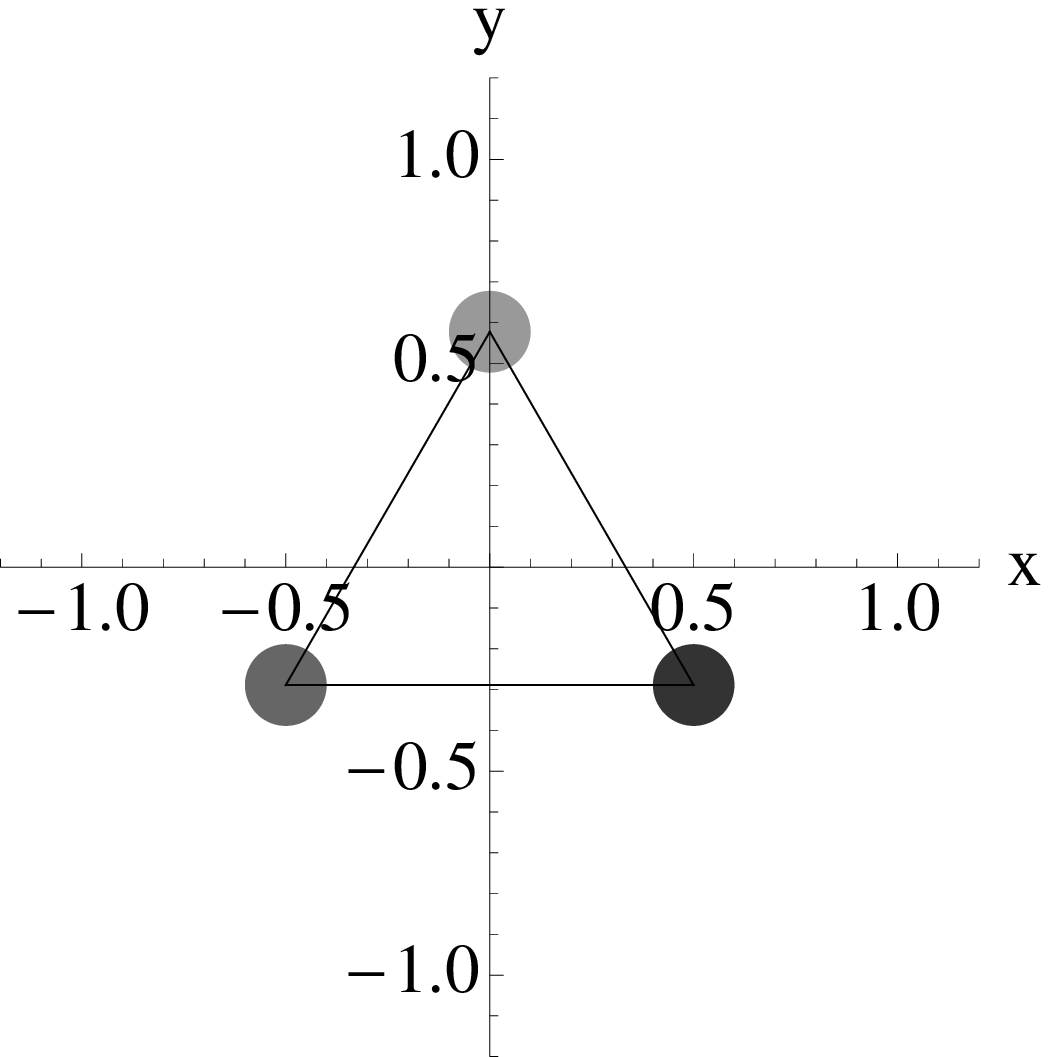}} &
\resizebox{50mm}{!}{\includegraphics{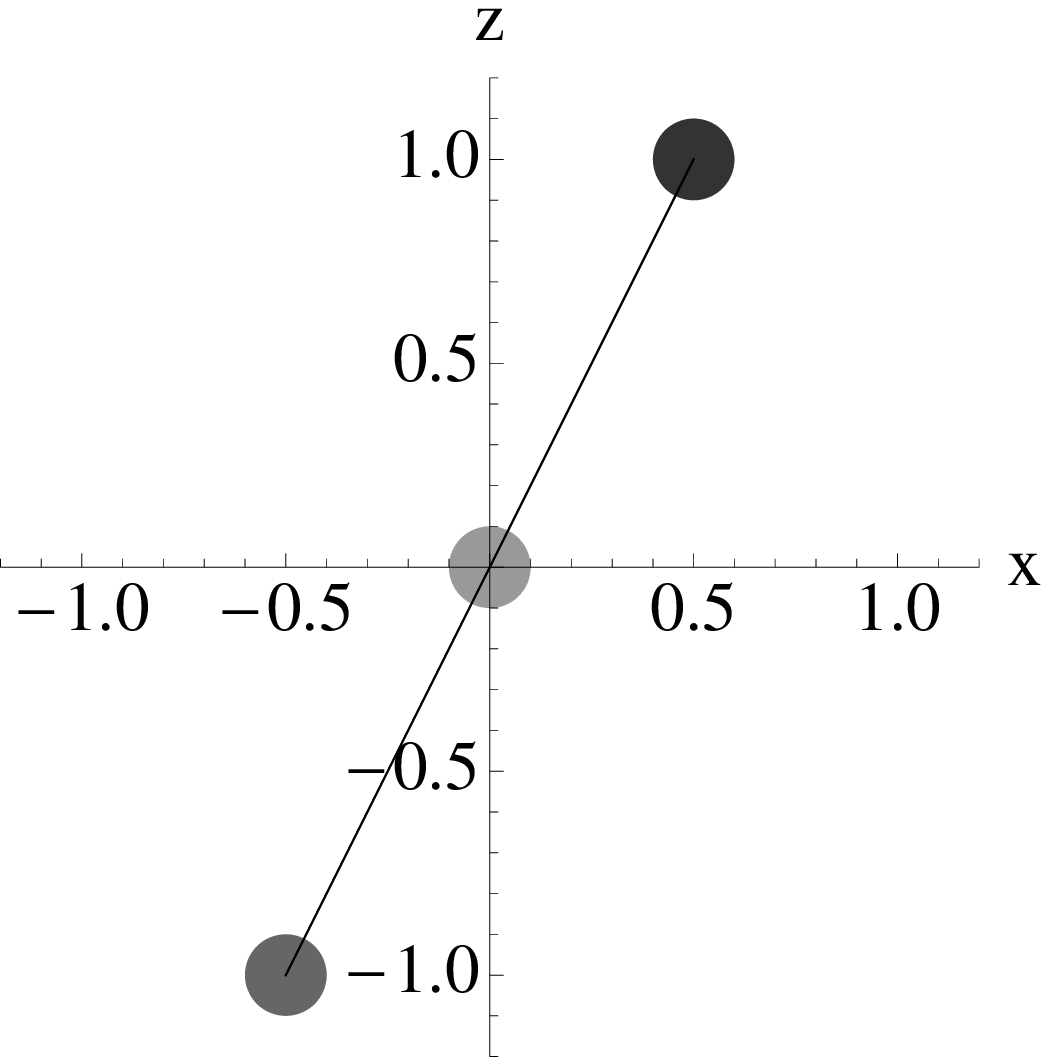}}
\end{tabular}
\caption{\label{baseShape}The stokeslet configuration used to check the scaling of $\lambda$ with the size of the object. To the left is a perspective view from an arbitrary direction. The pulling direction is toward the bottom of the cube, in the $-z$ direction. The center and right views show projections of the stokslets onto the $xy$ and $xz$ planes, respectively.}
\end{figure*}

To begin, we check the dependence of $\lambda$ on the size of the object. We do this by fixing a configuration of stokeslets and then computing $\lambda$ as we uniformly change the inter-stokeslet distances. The particular configuration we use is shown in Figure~\ref{baseShape}. It has the three stokeslets arranged so that their projection in the $xy$ plane is an equilateral triangle with side length $R$ centered about the origin, and their positions along the $\hat z$ axis are $0$ and $\pm R$. In this case, it does not matter which corner of the triangle is at which $z$ value; by symmetry, rearranging them can at most change the sign of $\lambda$, while its magnitude is our concern here. We will then move the forcing point to $R_p\hat z$, with $R_p\gg R$, and compute $\lambda$.

\begin{figure*}
\begin{tabular}{ccc}
\resizebox{50mm}{!}{\includegraphics{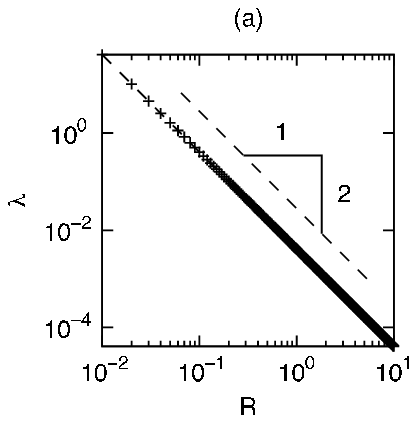}} &
\resizebox{50mm}{!}{\includegraphics{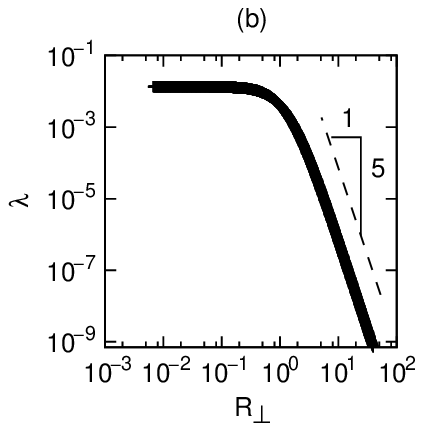}} &
\resizebox{50mm}{!}{\includegraphics{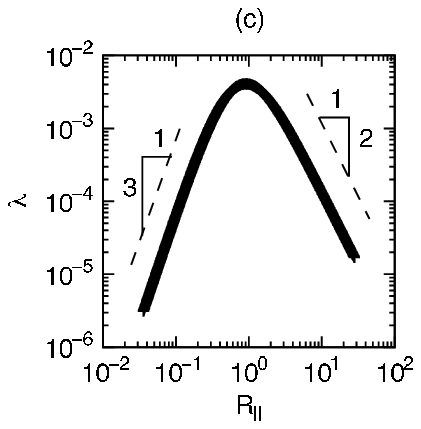}}
\end{tabular}
\caption{\label{scaling}Numerical results showing the scaling of $\lambda$ with $R$ (a), $R_\perp$ (b), and $R_\parallel$ (c), for configurations like that of Figure~\ref{baseShape}. In (a) there is a single scaling exponent of -2. In (b), for $R_{\perp}\gg R_{\parallel}=1$, we have a scaling exponent of -5, and for $R_\perp\ll R_{\parallel}=1$, it is constant. In (c), we have an exponent of 3 for $R_\parallel\ll R_{\perp}=1$, and -2 for $R_\parallel\gg R_{\perp}=1$.}
\end{figure*}

In Section~\ref{sec:propmat} we noted that the propulsion matrix depends linearly on $\eta$, so $\lambda\sim\eta^{-1}$. We can ignore this simple dependence on viscosity by setting $\eta = 1$. We will also set $\gamma = 6 \pi\eta\rho = 10^{-2}$, with $\rho$ in the same arbitrary distance units we use to measure $R$. As long as $\rho\ll R$, this is within the regime of small $\gamma$, so the precise value does not matter. 

As discussed in Section~\ref{sec:propmat}, all elements of $\mathbb T$ scale as an inverse length squared, so $\lambda$ must as well. Since $\lambda$ is independent of the Stokes radius and distance to the forcing point, as shown in Sections \ref{sec:nearlyfreedrain:propmat} and \ref{sec:chirality:distant}, we must form this length scale from the inter-stokeslet distances. Indeed, we can verify numerically that $\lambda\sim R^{-2}$, as shown in Figure \ref{scaling}(a). Since $R_g$ scales with $R$, it is clear that $\lambda\sim R_g^{-2}$. We can further try to break this dependence down into one based on $R_\perp$ and $R_\parallel$. To begin, we fix the $z$ positions of the stokeslets to be $0$ and $\pm 1$, and then vary the side length of the equilateral triangle. As shown in Figure \ref{scaling}(b), when the side length is long compared to the $z$ positions, we get $\lambda\sim R_\perp^{-5}$. This corresponds to a flat transverse object. For small side lengths, we get $\lambda\sim R_{\perp}^0$. This corresponds to an object that is elongated along the pulling direction. We can also see what happens when we fix the side length of the equilateral triangle in the transverse projection at 1, and instead vary the $z$ distance between stokeslets, putting them at $0$ and $\pm R_\parallel$. The results are shown in Figure \ref{scaling}(c). We see that for $R_\parallel \ll 1$, we get $\lambda\sim R_\parallel^3$, and for $R_\parallel\gg 1$, we get $\lambda\sim R_\parallel^{-2}$. 

Taken together, these observations suggest that we can write $\lambda = R_g^{-2} f(R_\parallel / R_\perp)$, where
\begin{equation}
f(x)\sim\left\{\begin{array}{cc}x^3; & x\ll 1\\ x^0 & x\gg 1.\end{array}\right.
\end{equation}
We can see that when $R_\parallel \gg R_\perp$, the function $f$ is a constant. Thus in this regime we know the scaling of $\lambda$ based on relative transverse and longitudinal sizes, and can focus on other aspects of the object's shape.

We will consider two general distortions of our shape from the previous one: first, we will relax the requirement that the $z$ values be equally spaced, in order to see the effect of bunching a pair of stokeslets together. Next, we will remove any restrictions on the transverse shape.

To characterize the bunching, we will use the inverse squared moment $Z$, defined by
\[
Z^{-2} = \frac 1n\sum_{\alpha=1}^n(r_z^\alpha-r_z^c)^{-2}
\]
This length $Z$ is dominated by the closest pairs of stokeslets. If we consider the ratio $Z/R_\parallel$, we get a dimensionless quantity which becomes large if some stokeslets are bunched close together.

\begin{figure*}
\resizebox{100mm}{!}{\includegraphics{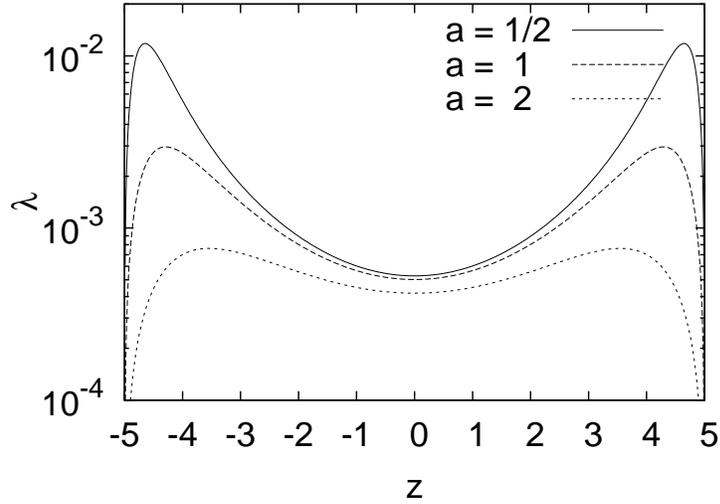}}
\caption{\label{bunchiness}The chiral coefficient $\lambda$ for a three-stokeslet object whose transverse projection is an equilateral triangle of side length $a$. Two of the longitudinal coordinates are fixed at $\pm 5$, and the third is varied over $z$ values between them.}
\end{figure*}

If a pair of stokeslets is bunched together, the hydrodynamic interactions between them become stronger. We expect this greater interaction to promote chiral behavior. Indeed, our numerical studies indicate that uneven spacing leads to larger $\lambda$. We again fix the transverse projection of the stokeslets to be an equilateral triangle, with side length $a$. We then choose the two extremal longitudinal projections to be at $\pm 5$, and allow the middle stokeslet position $z$ to vary between the other two. Figure \ref{bunchiness} shows the chirality as a function of $z$, for three different values of $a$. We see that there is a peak in $\lambda$ as the stokeslets approach each other, but it falls off if they get too close. The maximal $\lambda$ occurs when the longitudinal spacing is about equal to the transverse spacing.

To characterize the shape of the transverse projections, we consider the eccentricity of the inertia ellipse. If we define a projected tensor of inertia by
\[
\mathbb I_{ij} = \frac 1n\sum_{\alpha=1}^n(r_i^\alpha-r_i^c)(r_j^\alpha-r_j^c)
\]
for $i,j\in\{x,y\}$, then we can use
\[
\Delta = \frac{4\textrm{Det}(\mathbb I)}{\textrm{Tr}^2(\mathbb I)}
\]
as a measure of the eccentricity. It goes to zero when the stokeslets are collinear, and one when they are isotropically arranged.

We can now consider $\lambda$ as a function of both $\Delta$ and $Z/R_\parallel$. We confine ourselves to shapes with $R_p\gg R_\parallel\gg R_\perp$, which gives maximal $\lambda$ as seen above.

To see the dependencies, we generated $10^4$ random 3-stokeslet configurations, choosing each stokeslet from the box $[-1/2,1/2]\times[-1/2,1/2]\times[-10,10]$, and discarding it if $R_\parallel < 10 R_\perp$. Again we removed $\gamma$ and $\eta$ dependencies by taking $\gamma = 10^{-2}$ and $\eta = 1$.

The observed $\lambda$ values varied widely and irregularly. However, if we define the $p$th moment of the stokeslet positions
\[
U_p=\Bigg[\frac{2}{n(n-1)}\sum_{\alpha=1}^n\sum_{\beta=\alpha+1}^n|\vec r^\alpha-\vec r^\beta|^p\Bigg]^{1/p}
\]
and instead plot $\lambda (U_{-2})^2(Z/R_\parallel)^2$, we get a relatively smooth bounded function. Thus we can write
\begin{equation}
\lambda=(U_{-2})^{-2}\Big(\frac{Z}{R_\parallel}\Big)^2g(\{\vec r^\alpha\})
\end{equation}
where $g$ is a bounded function of its arguments.  

\begin{figure*}
\resizebox{100mm}{!}{\includegraphics{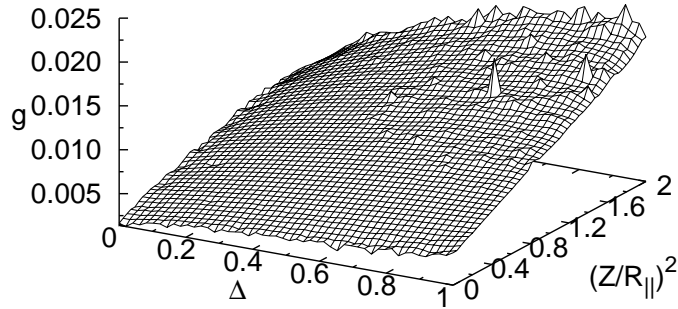}}
\caption{\label{gplot}The $g$ function plotted versus $\Delta$ and $(Z/R_\parallel)^2$. We can see that it is bounded and prefers high $\Delta$ and $Z/R_\parallel$.}
\end{figure*}

Figure \ref{gplot} shows a plot of $g$ as a function of $\Delta$ and $(Z/R_\parallel)^2$. From this plot, we can see a definite dependence on $\Delta$, indicating that $g$, and thus $\lambda$, prefer higher $\Delta$. This means that faster rotation occurs when the transverse projection is isotropic rather than elongated, while the object as a whole is long and slender.

In general, studying this simple 3-stokeslet case in the limit of distant forcing points has shown that the preferred shape for high chirality is a long and slender object. Along the length of the object, some clustering of stokeslets is preferred, and in the transverse plane it is beneficial to have an isotropic arrangements of stokeslets. 

So far we have only considered the magnitude of the chiral response for our three stokeslet systems. It would be convenient if there was an easy way to determine the sign of the chirality as well. We propose a method which seems to give acceptable results for those systems with large values of $|\lambda|$. 

We first order the stokeslets according to their longitudinal proximity to the forcing point. In the transverse projection, the ordering will form either a clockwise or counterclockwise triangle. We propose that these respectively correspond to a negative and a positive chirality. The physical argument for this triangle rule is that as the object sinks, the first stokeslet will have a stronger interaction with the second than the third, and so on. This will cause a slipstreaming effect, where the fluid behind the first causes less drag on the second behind it. This preferentially allows the object to move in that direction, much like a corkscrew.

To test this numerically, we generated $10^4$ triples of stokeslets chosen at random from the box $[-2,2]\times [-2,2]\times [-2,2]$. For each object we computed the chirality and applied the above triangle rule. The results are shown in Table~\ref{signGuess}. The triangle rule predicted the correct chirality roughly three quarters of the time. We anticipate as well that more chiral objects will be more likely to follow our sign convention, as the slipstreaming effect will be stronger. To test this, we repeat our comparison using only those configurations whose $|\lambda|$ value was larger than the average. As shown in Table~\ref{signGuess}, our method was indeed more accurate with the more chiral configurations. We can test this in another way by limiting ourselves to the more chiral configurations which we know arise when our three stokeslets are instead chosen from the box $[-1/2,1/2]\times [-1/2,1/2]\times [-10,10]$. In the case of these slender configurations, our method is quite effective. While it is not perfect, it can provide a reasonable guess at the sign.

\begin{table}
\begin{center}
\begin{tabular}{|l|c|c|c|}
\hline
& All configurations & Very chiral configurations & Slender configurations\\
\hline
Matching signs & 7518 & 2498 & 9868\\
Different signs & 2482 & 264 & 162\\
Percent matched & 75.2 & 90.4 & 98.7\\
\hline
\end{tabular}
\end{center}
\caption[Comparison of different methods for estimating the chiral sign]{Comparison between the number of times the actual sign of the chirality matched the sign estimated using our triangle rule. The very chiral configurations were selected from the rest via the criterion that their chirality be larger than the average. In addition, $10^4$ configurations were chosen at random from the slender regime studied earlier, where we expect to find the most chiral configurations.}
\label{signGuess}
\end{table}

Our explicit calculations above focused on the simplest stokeslet object that can have chirality: three stokeslets with the forcing point at infinity. We noted that such an object has five relevant degrees of freedom, but studied the effect of only four of them. To specify the minimal object completely therefore requires an additional parameter. One choice is to use the full $3\times 3$ inertial tensor, instead of its transverse projection. The principal axes of this tensor need not be aligned with the forcing direction, so we can take our additional parameter to be the smallest angle between a principal axis and the forcing direction. Evidently for the elongated objects with large $\lambda$ we have been studying, this angle is small, and does not have a major effect in this regime.

\section{Examples of behavior}
\label{sec:example}

\subsection{Numerical results}
\label{sec:example:numerical}

\begin{figure*}
\begin{tabular}{ccc}
\resizebox{50mm}{!}{\includegraphics{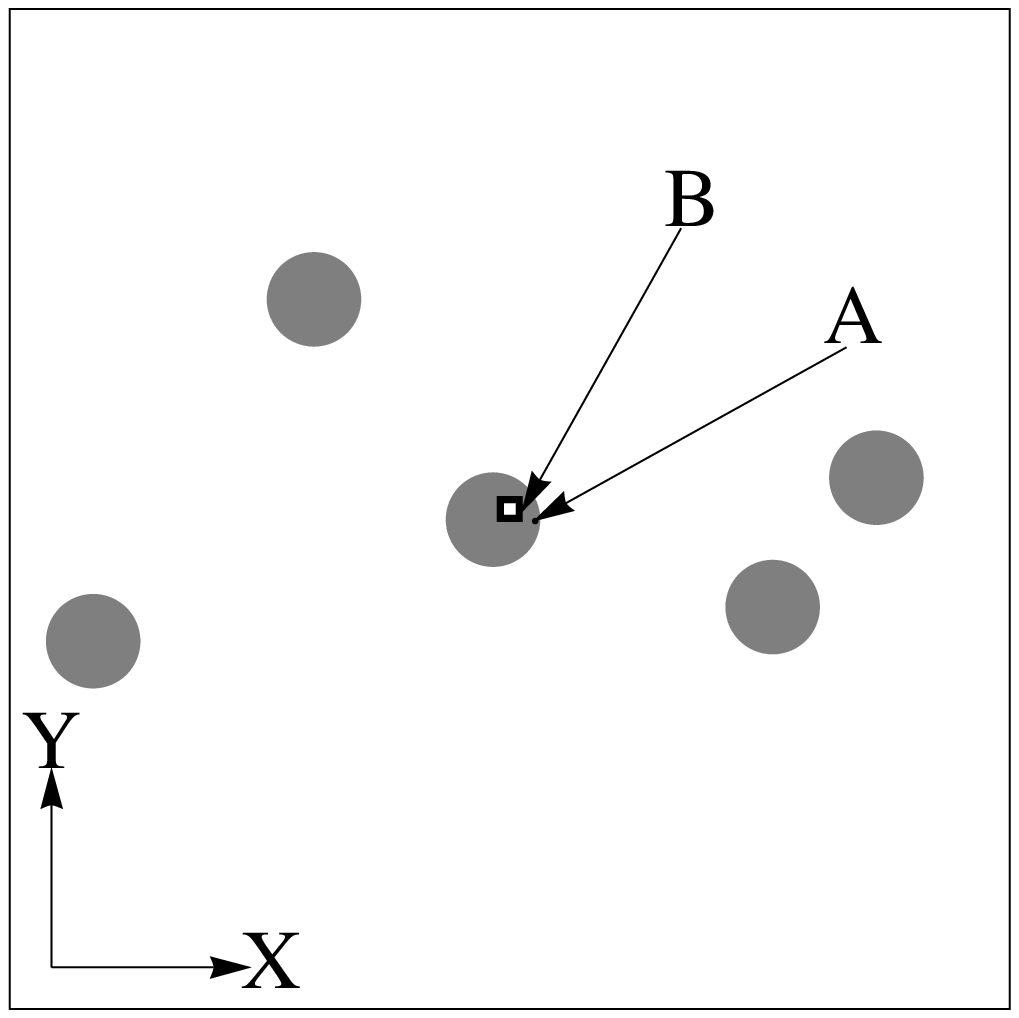}} &
\resizebox{50mm}{!}{\includegraphics{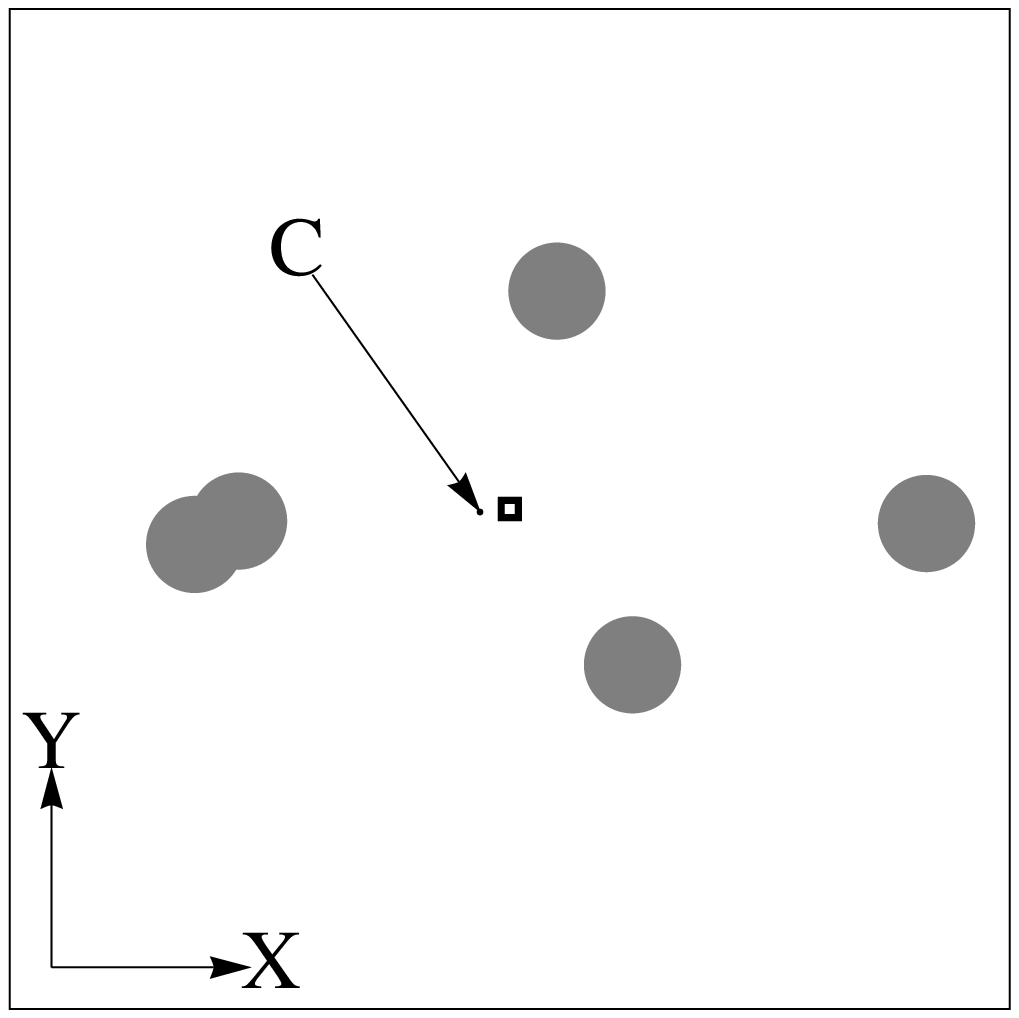}} &
\resizebox{50mm}{!}{\includegraphics{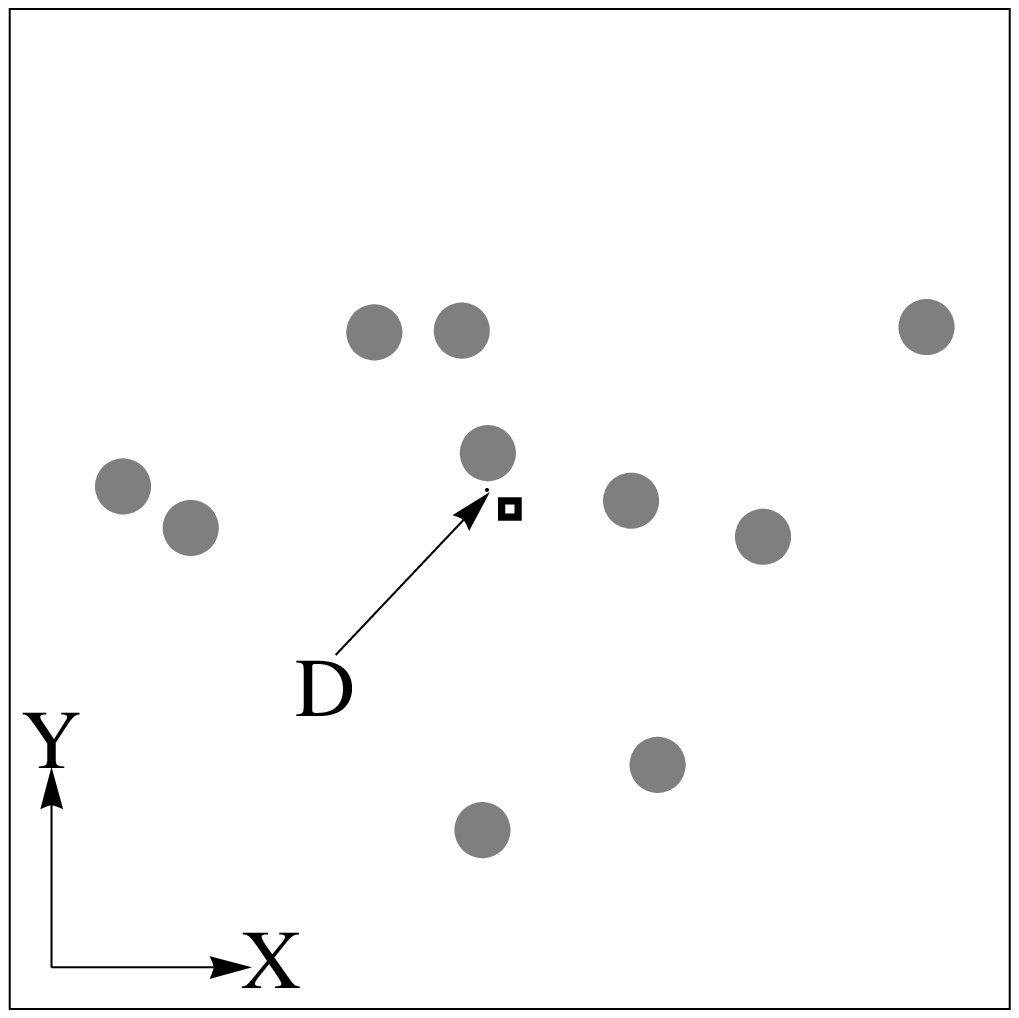}} \\
\resizebox{50mm}{!}{\includegraphics{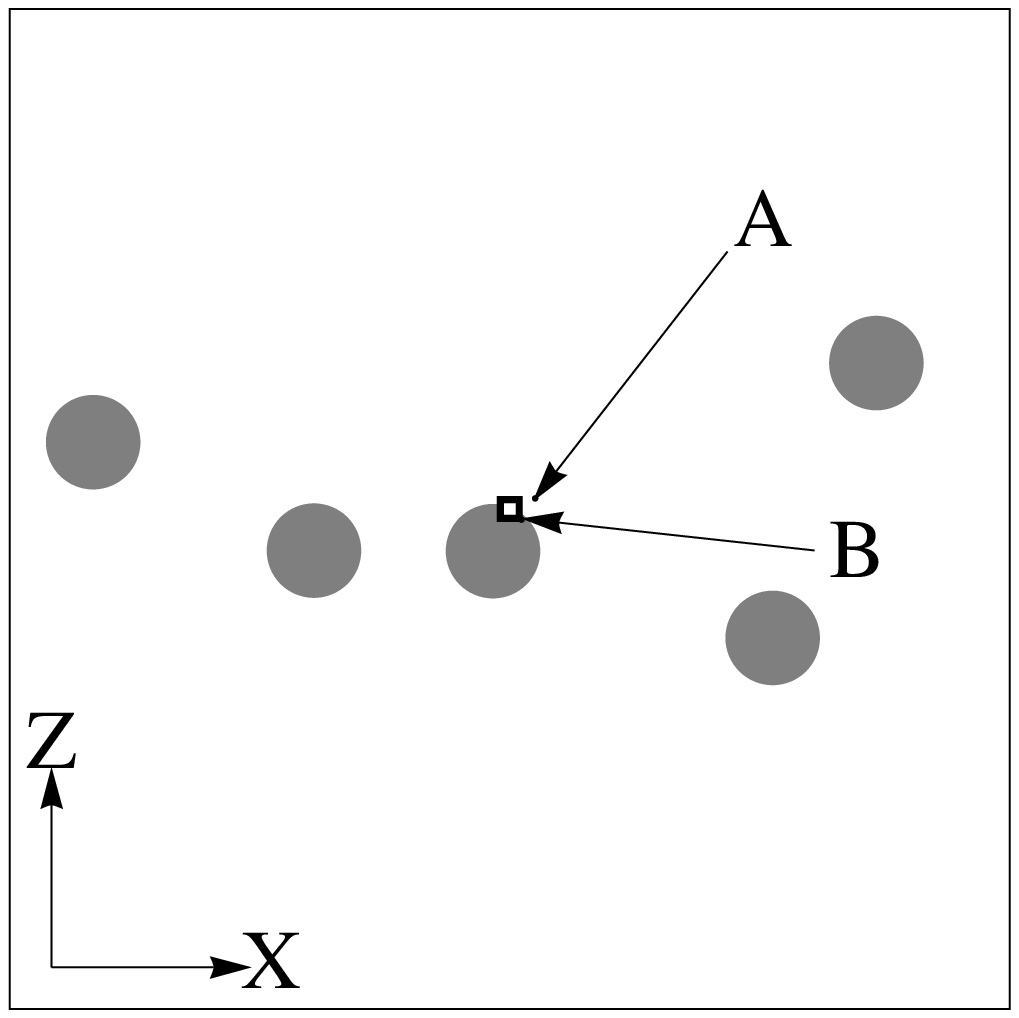}} &
\resizebox{50mm}{!}{\includegraphics{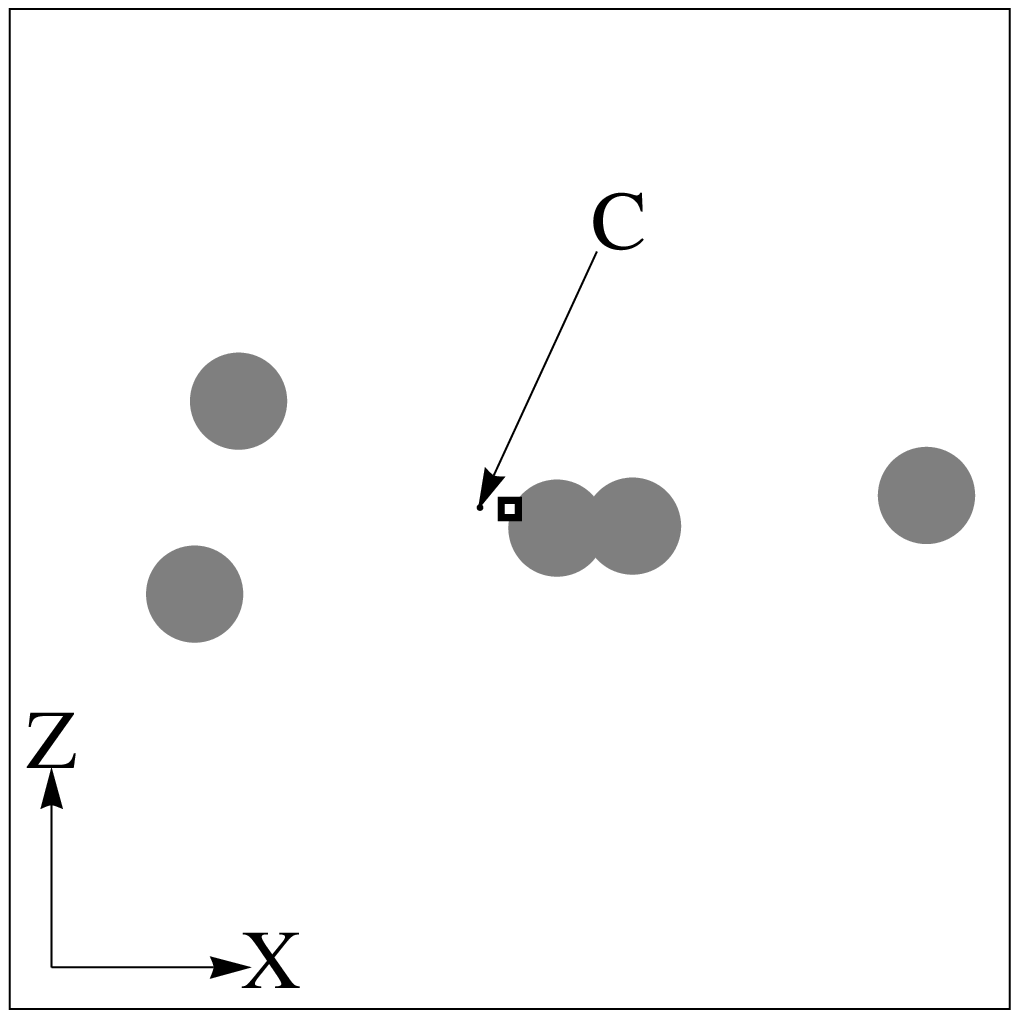}} &
\resizebox{50mm}{!}{\includegraphics{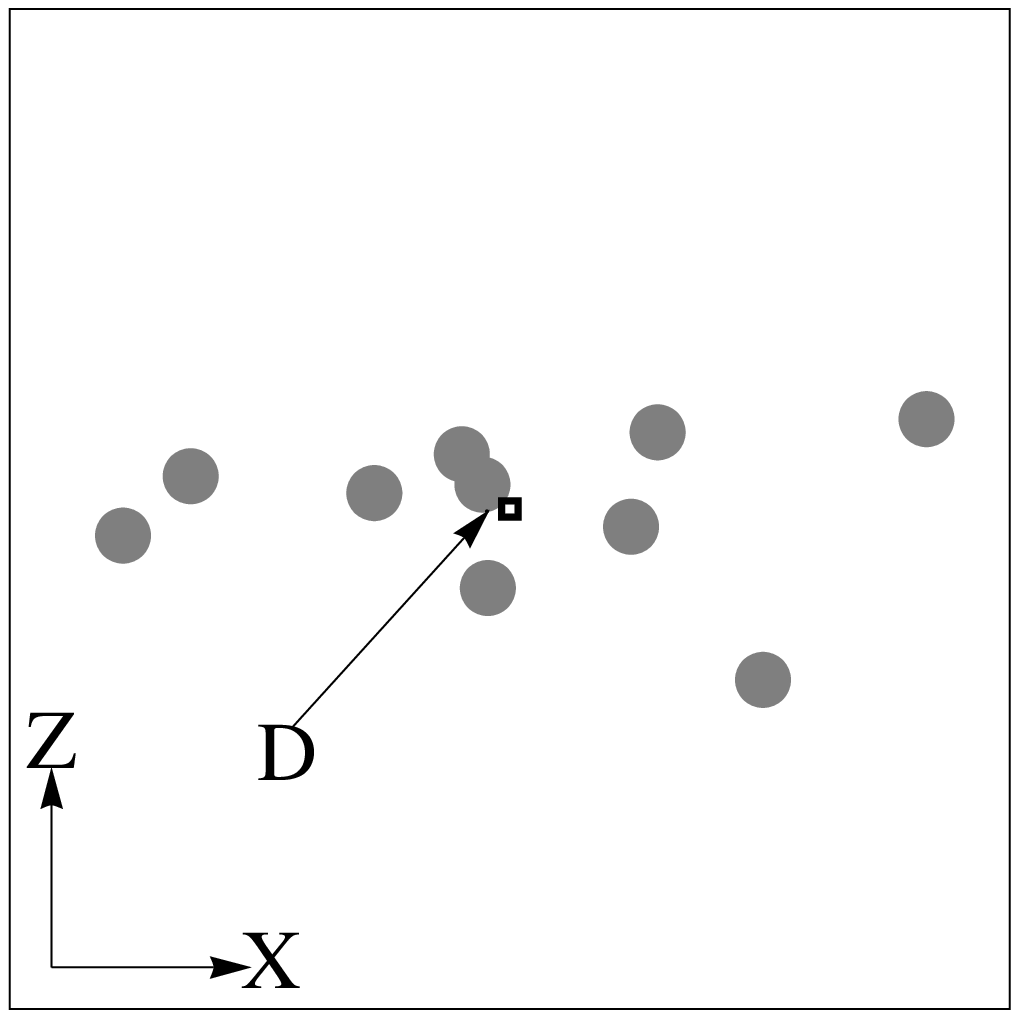}}
\end{tabular}
\caption{\label{numericalShapes} Projections of the four stokeslet configurations used in Section~\ref{sec:example:numerical}. In each image, the grey circles represent stokeslets, a small square marks the average stokelet position, and arrows point to the forcing points used. Objects A and B are identical except for the positions of their forcing points, and are shown on the left. Objects C and D are shown in the middle and on the right. Each object has been rotated so that the coordinate axes are aligned with the principal axes of the inertia tensor, with $\hat z$ and $\hat x$ corresponding to the largest and smallest of these, respectively. The size of the grey circles corresponds to the largest the stokeslets can be without causing the object to enter the tumble zone. In the case of the leftmost images, this is done with respect to object B.}
\end{figure*}

As a simple test of the results from Section~\ref{sec:nearlyfreedrain}, we can generate several stokeslet configurations at random, and verify that in the nearly free draining limit we get the simple chiral sedimentation predicted above, with the expected axis of rotation and angular velocity. We will do this with four objects: For object A, we form a five stokeslet object by picking random positions in the box $[-2,2]^3$ and setting the origin as the forcing point. Object B is the same as object A except for the location of the forcing point. This point is moved closer to the center of twisting in order to increase the tendency to tumble. Specifically, the center of twisting is determined at a particular choice of stokeslet radius, namely $2/3$ of the radius $\rho_{\textrm{max}}$ which would create contact between stokeslet spheres. The forcing point is then placed at this center of twisting, and remains there as the stokeslet radius is varied and the resulting motion measured. Object C is created the same was as object B, but with a different random choice of stokeslet positions. Finally, object D is a random ten stokeslet object, again with the origin moved as above. These are shown in Figure~\ref{numericalShapes}.

For each object, we first determine the axis we expect the object to rotate around in the nearly free draining limit. This is easy: as described in Section~\ref{sec:nearlyfreedrain:propmat}, the real eigenvector $\vec\lambda$ of the twist matrix is just the vector from the forcing point to the average stokeslet position, $\vec\lambda\equiv\vec r^c$. 

To find the angular velocity, we compute $\mathbb T_0$ and $\mathbb T_1$ as in Section~\ref{sec:nearlyfreedrain}, and form the basis $\vec v_1$, $\vec v_2$, and $\vec \lambda$ which puts $\mathbb T_0$ into Jordan canonical form. Let $\vec\lambda^T_d$ be the dual of $\vec\lambda$, which satisfies $\vec\lambda_d^T\vec\lambda=1$ and $\vec\lambda_d^T\vec v_i=0$. Then the real eigenvalue of $\mathbb T$ is just $\lambda = \vec\lambda_d^T\mathbb T_1\vec\lambda$, and we can find the angular velocity from $\omega=\lambda |F|$ with $|F|$ the magnitude of the sedimenting force. 

\begin{figure*}
\resizebox{150mm}{!}{\includegraphics{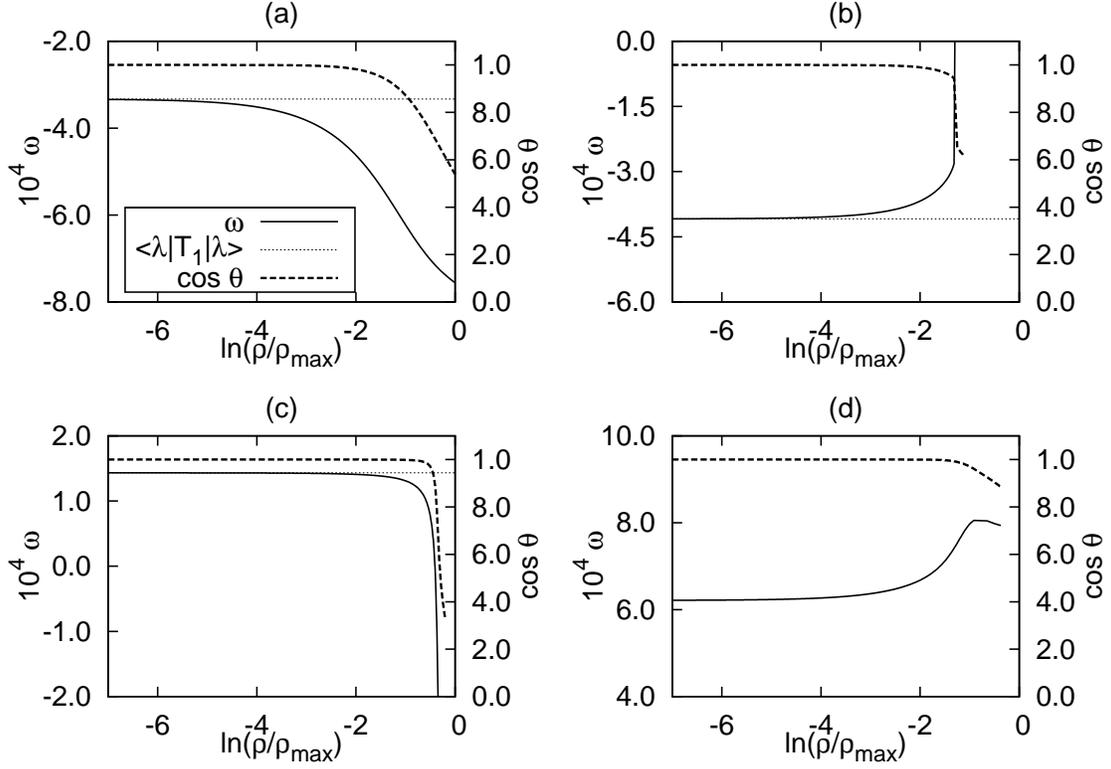}}
\caption{\label{testLimits} A comparison between the predicted values of the axis of rotation and angular velocity in the nearly free draining limit with the full results valid for all values of the drag coefficient. On the left vertical axis, the solid line represents the full $\omega$ when the object is undergoing the globally stable chiral motion, and the dotted line represents the value computed from the perturbative expressions in Section~\ref{sec:nearlyfreedrain}. On the right vertical axis is the cosine of the angle $\theta$ between the axis of rotation and the axis of rotation computed in the nearly free draining limit. (a), (b), (c), and (d) correspond to objects A, B, C, and D, respectively. In the case of (a), the object does not enter the tumble zone at all; before this happens, $\rho$ has increased to the unphysical point where the stokeslets overlap. However, the rest of the objects had their forcing points chosen in a manner which required them to be in the tumble zone for larger values of $\rho$, and their plots break off before $\rho_{max}$ is reached. }
\end{figure*}

In Figure~\ref{testLimits}, we compare these nearly free draining results with the results obtained from inverting Equation~\ref{eqn:propmatfromstokeslets}, using $|F|=\eta=1$. We see that the nearly free draining results hold over several decades of $\gamma$ values. Significant deviations occur only when the object is near the tumble zone. In the tumble zone, there is no single value of $\omega$ or $\cos\theta$ which can be plotted. However, we can see that we need to be quite close to the center of twisting for this to occur; the global stability and predictions from the nearly free draining limit made earlier are quite robust in practice.

We next study the effect of initial orientation on the sedimenting path, as Makino and Doi did for their skew propeller shape \cite{Makino:2003p4}. We do this by taking $N=100$ objects, each shaped as the object plotted in Figure~\ref{testLimits}(a) above, but with different random initial orientations. We then release them from the same point $(x,y,z)=(0,0,0)$, and consider their positions as functions of time, ignoring interactions between different objects. We determine these positions from the velocities given by $\vec V = \mathbb A(t)\vec F$ and $\vec\omega = \mathbb T(t)\vec F$. $\mathbb A(0)$ and $\mathbb T(0)$ can be found from Equation~\ref{eqn:propmatfromstokeslets}, and their time evolution is governed by the differential equations given in Section~\ref{sec:propmat}. While the elements of the these matrix are coupled together, the equations have no singularities, so we use Mathematica's \texttt{NDSolve} function \cite{Wolfram:2003} to  numerically find the solutions and expect reasonable accuracy. We use $\eta=1$, $\gamma=10^{-2}$, and supply a force $\vec F=\hat z$. 

\begin{figure*}
\resizebox{60mm}{!}{\includegraphics{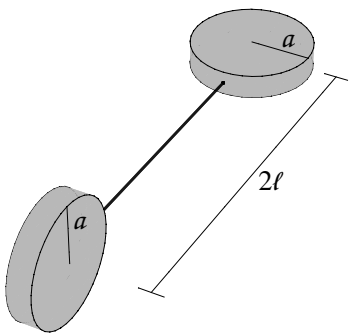}}
\caption{\label{propellerimage}The skew propeller shape used by \cite{Makino:2003p4}. The two orthogonal disks have radius $a$ and are fixed so their centers are a distance $2\ell$ apart. The center of twisting for this object coincides with the origin, so its twist matrix has three real eigenvalues.}
\end{figure*}

These results can be compared to those from the skew propeller shape, as well as a simple ellipsoid. The propeller consists of two orthogonal disks of radius $a$ attached via a thin rod so that their centers are a distance $2\ell$ apart, as shown in Figure~\ref{propellerimage}. The relevant portions of the mobility matrix are $\mathbb A = $diag$(a_x, a_x, a_z)$ and
\[
\mathbb T = \left(\begin{array}{ccc} 0 & b & 0\\ b & 0 & 0\\ 0 & 0 & 0 \end{array}\right)
\]
with
\begin{eqnarray*}
a_x & = & \frac{3 ( 4a^2 + 5\ell ^2)}{64 a \eta (5 a^2 + 6 \ell ^2)}\\
a_z & = & \frac{3}{64a\eta}\\
b & = &-\frac{3\ell}{64a\eta(5a^2+6\ell^2)}
\end{eqnarray*}
In the following, we use $\ell = 3a$, and then set $a=1$ in the same length units we used for our nearly free draining object. 

The skew propeller is an example of an object whose twist matrix is symmetric, and thus allows us to compare our nearly free draining object with something in the tumble zone. The ellipsoid allows a comparison with an object that has no translation - rotation coupling; its twist matrix is zero. We will choose its dimensions so that its alacrity matrix is the same as that of the skew propeller.

To do the comparison, we can look at the width of the distribution of particles as a function of time, as well as the spread in the $z$ direction:
\begin{eqnarray}
w(t) &=& \frac 1N \sum_{i=1}^N\sqrt{x_i^2(t)+y_i^2(t)}\\
h(t) &=& \frac 1N \sum_{i=1}^N | z_i(t) - \langle z(t)\rangle |
\end{eqnarray}
where $\{x_i(t),y_i(t),z_i(t)\}$ is the position of the particle at time $t$, and $\langle z(t)\rangle = \frac 1N\sum_iz_i(t)$ is the average $z$ position of the ensemble at time $t$.

\begin{figure*}
\begin{tabular}{cc}
\resizebox{80mm}{!}{\includegraphics{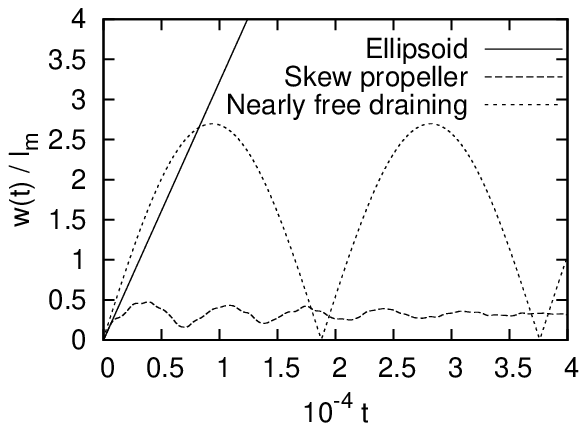}}
\resizebox{80mm}{!}{\includegraphics{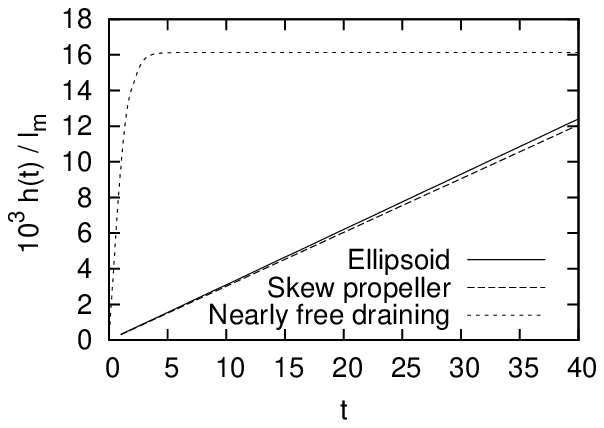}}
\end{tabular}
\caption{\label{spread}Plots of normalized $w(t)$ and $h(t)$ as functions of time. The width increases linearly with time for the ellipsoidal particle, but remains bounded for the particles with a nonzero twist matrix. The spread of the particles increases linearly with time for both the skew propellers and the ellipsoids, but after an initial transient remains constant for the nearly free draining particles.}
\end{figure*}

Figure~\ref{spread} shows $w$ and $h$, normalized by the maximum linear distance between two points on the object, $l_m$. The ellipsoids must distribute themselves on the surface of a sphere sinking at a constant velocity whose radius increases with constant velocity \cite{Makino:2003p4}. Thus $h$ and $w$ are both linear in time for ellipsoids. 

The widths of the distributions for the skew propellers and our sample object are both bounded. The skew propellers evidently approach a constant $w$, while the nearly free draining objects have a $w$ which oscillates at their rotation frequency $\omega$. The spread $h$ for the skew propellers in the tumble zone increases linearly. However, we see that after an initial transient motion, the longitudinal spread of our sample particles remains constant. This is because all of them begin to sediment in the same regular manner.

Thus, overall we see that the ellipsoids spread out into a spherical volume as they sink, with radius increasing linearly. The skew propellers spread out in a cylindrical shape parallel to the applied force. The length of the cylinder increases linearly, while the radius undergoes decaying oscillations about a value smaller than the linear extent of each object. The nearly free draining particles spread out over a flat disk with constant longitudinal spread and a radius which oscillates at the same frequency that each particle spins at. The amplitude of these oscillations is slightly larger than the maximum extent of the object.

\subsection{Experimental illustrations}
\label{sec:example:experimental}

In order to verify that the chiral rotation discussed above is significant in practice, we created some arbitrarily shaped bodies and observed their sedimentation. We used both viscous and non-viscous solvents. This allows us to gauge the importance of inertial effects.

For the viscous solvent, we cut small objects out of a rod of nylon plastic, a few millimeters in length in their longest direction. We also took small lengths of copper wire and bent them into twists or knots. Our objects were allowed to sediment in a 700 mL beaker filled with vegetable oil. Such oils have kinematic viscosities of the order 30 cSt \cite{Bantchev:2008p1324}, and our nylon pieces fell at around 0.2 cm/s, giving a Reynolds number of slightly less than 1, well within the Stokes regime. The copper twists, which fell more quickly, are still at Reynolds numbers where inertial effects are not expected to be important. 

\begin{figure*}
\resizebox{60mm}{!}{\includegraphics{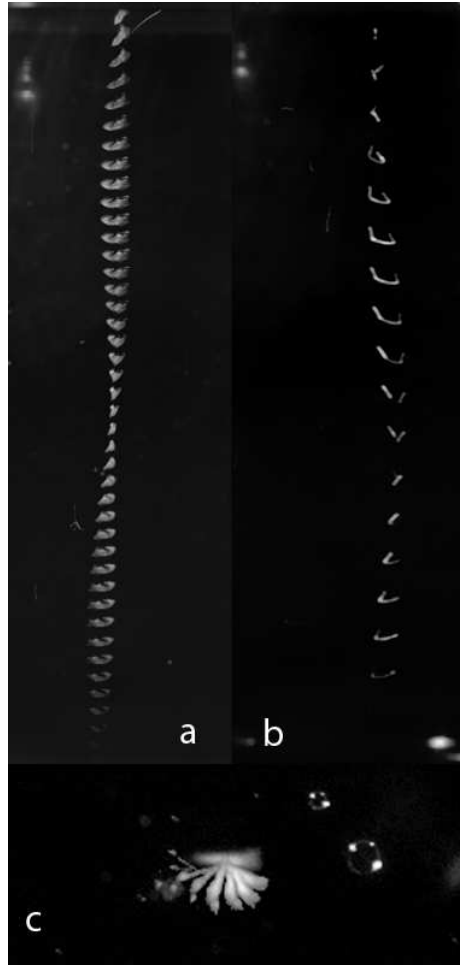}}
\caption{\label{KeimKrapf}(a): A multiple-exposure image of an irregular piece of nylon sedimenting in vegetable oil. The nylon piece is a few millimeters in length, and the pictures were taken about 1 second apart. It is clearly rotating around the vertical axis. (b): A multiple-exposure image of a fine piece of copper wire sedimenting in vegetable oil. The pictures were taken about 0.3 seconds apart, and the distance scale is the same as in (a). The object is rotating about the vertical axis as it follows a helical path down. (c): A multiple-exposure image of the same piece of nylon in (a), though not at the same time. From above, the helical path is more apparent. More images and movies are available at \imageurl}
\end{figure*}

We used tweezers to hold each object just below the surface, then released it and used a camera to take pictures at a rate of about 3 frames per second. For these uniform materials, the forcing point is the center of mass, which we expect to be close to the center of reaction. Thus it is not clear from our arguments above that these objects should be outside the tumble zone. Nevertheless, we were able to see chiral sedimentation with many of these objects. Figures~\ref{KeimKrapf}(a,b) show multiple-exposure views for both a nylon piece and a twist of fine copper wire. The helical path is obvious for the copper piece, but less so for the nylon. Figure~\ref{KeimKrapf}(c) shows the same nylon piece, in a separate run, from above. Here the helical nature of the path is easier to see.

In addition to the objects shown, we tested over a dozen other objects made in the same way. Some displayed little or no rotation, and simply settled into a preferred orientation. Some of the heavier ones, which sank very rapidly, showed a slight rotation about axes other than the vertical. We cannot tell if this was actually a case of the objects tumbling; we suspect that it was instead an initial reorientation which aborted when they hit the bottom of the beaker before reaching their preferred orientation.

For comparison, we dropped small shards of brittle plastic into a salt water solution, whose viscosity was lower than the oil's. The objects, cut from a disposable spoon, were a few millimeters in size. Salt was added to the water to achieve nearly neutral buoyancy without greatly affecting the viscosity. In this solution, the objects fell at around 1 cm/s, giving a Reynolds number $\lesssim 100$, which is not fully in the regime of Stokes flows. However, we still observed chiral sedimentation, so even at this Reynolds number the inertial effects do not appear to change the motion qualitatively.

In these studies, we monitored for residual circulation in the water by putting a small cylinder of floating plastic on the surface. This cylinder remained stationary, indicating that any residual flow is much smaller than the chiral motions. 

\begin{figure*}
\resizebox{50mm}{!}{\includegraphics{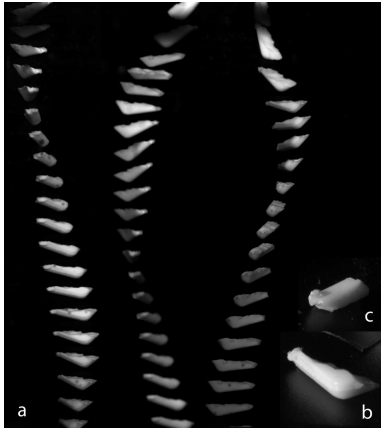}}
\caption{\label{KeimWitten}(a): Three multiple-exposure pictures of the 9 millimeter long object pictured in (b) as it sediments in salt water. The pictures were taken about 0.15 seconds apart. In each case the twist about the vertical axis as it moves in a helix is clearly visible, indicating that the chiral effects on sedimentation are similar to those in the viscous solvent of Figure~\ref{KeimKrapf}. Each picture corresponds to a different initial orientation of the object. Though the transient motion was different in each case, it always ended up turning to the same preferred orientation and twisting in the same direction. (c): Another object cut from a plastic spoon. This object has two stable orientations, which lead to twists about the vertical axis in opposite directions.}
\end{figure*}

With these plastic pieces, no ongoing tumbling motion was seen; either there was no rotation, or else they rapidly reoriented themselves and twisted around the vertical axis. Figure~\ref{KeimWitten}(a) shows a multiple-exposure picture of a typical path. This object, shown close-up in Figure~\ref{KeimWitten}(b), turned to the same preferred direction regardless of initial orientation, and always rotated with the same sign. However, this is not the only behavior; the object pictured in Figure~\ref{KeimWitten}(c) had two opposite orientations which were stable. These produced opposite signs for the rotations. In addition, some objects showed negligible rotation, though they did go to the same stable orientation. This must correspond to an instance where $\lambda\sim 0$. 

\section{Discussion}
\label{sec:discussion}

In the foregoing we have explored how slowly-sedimenting noncompact objects of generic shape rotate as they sink, revealing chiral structure. These objects were represented as collections of stokeslets, which are known to provide a good representation of a broad range of real objects \cite{Carrasco:1999p19}. We infered the propulsion matrix from the matrix of Oseen interactions between pairs of stokeslets. This propulsion matrix is sufficient to determine the entire motion under slow sedimentation at low Reynolds numbers \cite{Happel:1983}. To determine the chiral rotation, it is sufficient to know the $3\times 3$ twist matrix $\mathbb T$ derivable from the propulsion matrix. In the case when $\mathbb T$ has only one real eigenvalue, there is globally stable motion corresponding to rotation about the corresponding eigenvector \cite{Gonzalez:2004p10}.

Though all chiral rotation must vanish when there are no hydrodynamic interactions, in the nearly free draining limit where these interactions are arbitrarily small, there is nevertheless a constant and finite rotation about a fixed axis. The angular velocity in this limit is independent of the strength of the interactions, and the rotation axis approaches the line between the forcing point and the center of reaction.

The features of an object that determine its chiral sedimentation are unexpectedly subtle. Indeed, the rotation rate depends on the stokeslet positions in a singular way, with unevenly spaced stokeslets giving the largest response. For such configurations it is the nearest distance that dominates, and small displacements of the stokeslets on the order of this shortest distance suffice to reverse the sign of $\lambda$. Thus $\lambda$ is not a gross indicator of overall chiral shape. Instead, it is a local probe, sensitive to local orientations relative to the overall object. The maximum responses occurred for thin, screwlike objects. Similar objects at the microscopic scale include biological filaments such as f-actin or microtubules.

The connection between our simple stokeslet objects and real objects has not been fully explored in this paper. Carrasco and de la Torre \cite{Carrasco:1999p19}, for example, describe methods for implementing the stokeslet model which appear applicable to the objects we discuss. Thus, rather than predicting the chiral response of any real object, we focused instead on finding the scaling and analytical asymptotic behavior for nearly free draining objects. 

We have developed an empirical rule to predict the sign of the chirality for some simple objects. However, this method should be improved. We would like to find a simple method to determine the chiral sign that is not only more accurate, but will also generalize to arbitrary objects. We also would like to establish analytically the scaling that we empirically determined in Section~\ref{sec:chirality:shape}, and to include the effects of brownian motion.

The free draining limit we use is physically approachable for the sedimentation of some large molecules or other polymers, formed by assembling macromolecules or colloidal particles. One could conceive of attaching a fluorescing group to such a molecule, and then using fluorescence polarization in a centrifuge to measure the spinning rate. The spinning rate could be used to characterize the object. 

Even in cases where the objects are not nearly free draining, we expect most of our conclusions to apply qualitatively; the nearly free draining limit is not the only way to escape the tumble zone, and general objects without symmetry will often see the globally stable behavior.

The chiral sedimentation treated here is only one example of how a colloidal object of irregular shape might respond in a chiral way. For example, objects sedimenting in shear flows can undergo net lateral drift according to chirality \cite{Doi:2005p20}. Varying sedimenting forces periodically in time could also be used to probe further properties of the propulsion matrix. Molecules of sub-micron scale such as folded RNA must also exhibit chiral sedimentation, though they will be greatly influenced by thermal brownian motion. Beyond the context of hydrodynamics, such objects can show chirality via their self-assembly properties. For example, two copies of a chiral globular protein have a most favorable orientation for binding. When many such copies self-assemble in this way, the least constraining mode of assembly is a one dimensional stack. Such a stack must in general show a chiral twist which may limit the stack's potential to stick to its neighbors. Aggeli et. al. \cite{Aggeli:2001p11857} use this as a model for the formation of peptide fibrils. This generic view may account for the prevalence of one-dimensional assemblies of biological molecules. Such responses are a promising course of study for the future. 

\section{Conclusion}
\label{sec:conclusion}

The most classic chiral response of microscopic matter, the rotation of the polarization of light, has been studied for over a century. Here we have discussed an equally fundamental response: the chiral interaction of an irregular object with a surrounding viscous liquid. In this case the chiral properties arise entirely from the object's geometry. We have seen that macroscopic objects of arbitrary shape have readily observed chiral sedimentation. The greatest response seems to occur when the drag is concentrated at one end of an elongated object. This study is only a first step towards understanding how shape creates chiral responses in colloid-scale materials. There are numerous ways to explore various shapes and numerous other responses, as sketched above. Understanding how shape determines chiral response should be valuable as a way of assessing the shapes of unknown objects and as a way of designing shapes to create desired responses. 

\begin{acknowledgments}
The authors are grateful to Phillipe Cluzel, Jean-Francois Joanny, Konstantin Turitsyn, and Sidney Nagel for useful discussions.  This work was supported in part by the National Science Foundation's MRSEC Program under Award Number DMR-0213745, and the Department of Education's GAANN Fellowship.\end{acknowledgments}

\bibliography{/Users/krapfn/Documents/Papers/bib_2008_07_29,/Users/krapfn/Documents/Papers/bookbib}

\end{document}